\documentclass[showpacs,preprintnumbers,amsmath,amssymb,APSl,prd,nofootinbib,superscriptaddress,10pt]{revtex4-2}
\usepackage{graphicx}
\usepackage{caption}
\usepackage{subcaption}
\usepackage{xcolor}
\usepackage{hyperref}
\usepackage{amsfonts}
\usepackage{bm}
\usepackage[a4paper, margin=1.2cm]{geometry}
\usepackage{mathrsfs}
\usepackage{soul}
\usepackage{makecell,multirow}
\usepackage{rotating}
\usepackage[utf8]{inputenc}
\usepackage{amsmath}
\usepackage{makecell}

\usepackage{amssymb}
\usepackage{tensor}
\usepackage{graphicx}
\usepackage{soul}
\usepackage[normalem]{ulem}
\usepackage[authormarkup=color]{changes}
\definechangesauthor[name={AS}, color=black]{as}
\usepackage{bigints}
\usepackage{bbm}
\usepackage{graphicx}
\usepackage{subcaption}
\usepackage{epsf}
\usepackage{bm}
\usepackage{amsmath}
\usepackage{amsfonts}
\usepackage{amssymb}
\usepackage{graphicx}
\usepackage{tabularx}
\usepackage{multirow}
\usepackage{color}%
\providecommand{\U}[1]{\protect\rule{.1in}{.1in}}

\newcommand{\ie}{\begin{equation}}
\newcommand{\fe}{\end{equation}}

\newcommand{\mincir}{\raise
-3.truept\hbox{\rlap{\hbox{$\sim$}}\raise4.truept\hbox{$<$}\ }}
\newcommand{\magcir}{\raise
-3.truept\hbox{\rlap{\hbox{$\sim$}}\raise4.truept\hbox{$>$}\ }}

\providecommand{\U}[1]{\protect\rule{.1in}{.1in}}

\usepackage{tikz,xcolor,hyperref}

\usepackage{hyperref}             
\hypersetup{
    colorlinks=true,              
    breaklinks=true,              
    citecolor=blue,               
    linkcolor=[rgb]{0,0.5,0.9},   
    urlcolor=red,                 
    filecolor=green               
}

\definecolor{lime}{HTML}{A6CE39}
\DeclareRobustCommand{\orcidicon}{%
	\begin{tikzpicture}
	\draw[lime, fill=lime] (0,0) 
	circle [radius=0.16] 
	node[white] {{\fontfamily{qag}\selectfont \tiny ID}};
	\draw[white, fill=white] (-0.0625,0.095) 
	circle [radius=0.007];
	\end{tikzpicture}
	\hspace{-2mm}
}

\foreach \x in {A, ..., Z}{%
	\expandafter\xdef\csname orcid\x\endcsname{\noexpand\href{https://orcid.org/\csname orcidauthor\x\endcsname}{\noexpand\orcidicon}}
}

\begin{document}

\title{\Large{Light propagation and gravitational lensing effects in charged Kalb–Ramond spacetime in nonlinear electrodynamics}}

\author{C. F. S. Pereira}
	\email{carlosfisica32@gmail.com}
	\affiliation{Departamento de F\'isica e Qu\'imica, Universidade Federal do Esp\'irito Santo, Av.Fernando Ferrari, 514, Goiabeiras, Vit\'oria, ES 29060-900, Brazil.}

    \author{Marcos V. de S. Silva}
	\email{marcos.sousa@uva.es}
\affiliation{Department of Theoretical Physics, Atomic and Optics, Campus Miguel Delibes, \\ University of Valladolid UVA, Paseo Bel\'en, 7,
47011 - Valladolid, Spain}
	\author{A. R. Soares}
    \email{adriano.soares@ifma.edu.br}
	\affiliation{Grupo de Investiga\c{c}\~{a}o em Educa\c{c}\~{a}o Matem\'{a}tica,  R. Dep. Gast\~ao Vieira, 1000, CEP 65393-000 Buriticupu, MA, Brazil.}


\author{A. A. Ara\'{u}jo Filho}
\email{dilto@fisica.ufc.br}
\affiliation{Departamento de Física, Universidade Federal da Paraíba, Caixa Postal 5008, 58051--970, João Pessoa, Paraíba,  Brazil.}
\affiliation{Departamento de Física, Universidade Federal de Campina Grande Caixa Postal 10071, 58429-900 Campina Grande, Paraíba, Brazil.}
\affiliation{Center for Theoretical Physics, Khazar University, 41 Mehseti Street, Baku, AZ-1096, Azerbaijan.}

\author{R. L. L. Vit\'oria}
    \email{ricardovitoria@professor.uema.br/ricardo-luis91@hotmail.com}
	\affiliation{Programa de Pós-Gradua\c c\~ao em Engenharia Aeroespacial, Universidade Estadual do Maranh\~ao, Cidade Universit\'aria Paulo VI, São Lu\'is 65055-310, MA, Brazil}
    \affiliation{Faculdade de F\'isica, Universidade Federal do Par\'a, Av. Augusto Corr\^ea, Guam\'a, Bel\'em, PA 66075-110, Brazil}

\author{H. Belich} 
\email{humberto.belich@ufes.br}
\affiliation{Departamento de F\'isica e Qu\'imica, Universidade Federal do Esp\'irito Santo, Av.Fernando Ferrari, 514, Goiabeiras, Vit\'oria, ES 29060-900, Brazil.}   


\begin{abstract}
In this work, we theoretically investigate the deflection of light for strong- and weak-field regimes in the background of an electrically charged BH described in Kalb-Ramond gravity, which introduces the Lorentz symmetry violation parameter $l$, as well as the control of the degree of nonlinearity incorporated by electrodynamics through the parameter $\gamma$. We analytically constructed the expansion coefficients in both limits and used them as a basis to investigate gravitational lensing effects through observables, taking into account the variation of the parameters involved in the model, both for the canonical field and the phantom case.

\end{abstract}
\maketitle

\tableofcontents


\section{Introduction}\label{sec1}

Singular black hole (BH) solutions arise naturally within the framework of general relativity (GR) as a result of the gravitational collapse process, being interpreted as the final stage of the stellar evolution of supermassive stars. The pioneering and most elementary theoretical formulation of these objects was presented by Karl Schwarzschild, consisting of a vacuum solution of the field equations established through Einstein's equations \cite{INTRO1,INTRO2,INTRO3,INTRO4}. These geometric structures generally have as their main ingredients the emergence of event horizons and singularities, and for this configuration, which is the simplest geometry, all other quantities are defined as functions only of the mass parameter. A somewhat richer structure emerges in the construction of the Reissner-Nordström geometry, which, in a simplified way, describes an electrically charged BH. In this case, we have an electromagnetic vacuum, and in addition to the presence of a singularity and an event horizon, a new structure called the Cauchy horizon appears. This geometry, as with the Schwarzschild BH, is spherically symmetric and static; however, now all other quantities can be expressed in terms of the parameters of the mass $M$ and the electric charge $Q$. These two geometries described above are extended to astrophysical scenarios through the stationary spacetimes of Kerr and Kerr–Newman \cite{INTRO1,INTRO2}.

The static BH configuration supported by a Kalb-Ramond field was first constructed in Ref.~\cite{yang2023static}. Since its introduction, this geometry has motivated a broad range of investigations. Studies have examined its quasinormal spectra \cite{araujo2024exploring}, transmission probabilities and greybody factors \cite{guo2024quasinormal}, quantum emission and evaporation processes \cite{AraujoFilho:2024ctw}, and gravitational deflection of light \cite{junior2024gravitational}. The orbital structure of test particles, including circular trajectories and quasi--periodic oscillations, has also been analyzed \cite{jumaniyozov2024circular}, as well as the accretion of collisionless Vlasov matter \cite{jiang2024accretion}.

Subsequent developments generalized the original setup. A non--commutative deformation of the Kalb-Ramond BH has been proposed \cite{AraujoFilho:2025jcu}, extending the framework to incorporate short--distance corrections. In addition, an electrically charged counterpart was formulated in Ref.~\cite{duan2024electrically}, followed by several works exploring its thermodynamic behavior, stability, and dynamical properties \cite{heidari2024impact,al-Badawi:2024pdx,aa2024antisymmetric,Zahid:2024ohn,chen2024thermal}.

Further refinements revealed that the original construction did not exhaust the full solution space. Additional branches were identified in Refs.~\cite{Liu:2024oas,Liu:2025fxj}, where previously overlooked configurations were systematically derived. These extended solutions have since been employed in different physical contexts, including analyses of entanglement degradation in curved backgrounds \cite{Liu:2024wpa} and applications to neutrino propagation phenomena \cite{Shi:2025xkd,Shi:2025rfq} for instance. Beyond these aspects, Kalb-Ramond gravity has been explored from several complementary perspectives. Investigations have addressed perturbative stability and phantom sectors \cite{Sekhmani:2025jbl}, topological BHs in anti--de Sitter spacetimes \cite{EslamPanah:2025oqy}, branching structures and non--extensive thermodynamics \cite{Sucu:2026nkw}, as well as optical characteristics and particle dynamics \cite{Ahmed:2026vfk}, among other topics \cite{AraujoFilho:2025fwd}.

More recently, a distinct class of Kalb-Ramond BHs sourced by ModMax nonlinear electrodynamics has been introduced \cite{Sekhmani:2025jbl}, with additional properties examined shortly thereafter \cite{Al-Badawi:2025ejf}. Despite this growing body of work, a treatment of gravitational lensing in this ModMax--sourced scenario has not yet been carried out. The present study addresses this omission by providing a detailed analysis of light deflection in the corresponding geometry.

In Section \ref{sec3} of this work, we present some general properties for the model as well as derive the conserved and geodesic quantities. In section \ref{sec4} we analyze optical effects for the weak- and strong-field regimes. In Section \ref{sec5} we apply the gravitational lensing technique to the model and construct the observables. Finally, in Section \ref{sec6}, we present the final considerations and conclusions.

\section{Kalb-Ramond-ModMax spacetime and geodesic equations}\label{sec3}

In this work, we will adopt the methodology used to calculate the deflection of light moving in a gravitational field and describing a spherically symmetric and static configuration for an electrically charged BH under the effects of Lorentz symmetry violation resulting from the coupling of nonlinear electrodynamics with Kalb-Ramond self-interaction fields, which was initially introduced by authors in Ref. \cite{Sekhmani:2025jbl}. This newly presented class of BH has already been explored in several scenarios, such as in the tidal force study where the authors investigate both radial and angular tidal forces on the event horizon, angular deviation, and the Roche limit \cite{Junior:2026upy}. These analyses were extended to particle dynamics, thermal properties, non-extensive thermodynamics, and observational signatures in Refs. \cite{Al-Badawi:2025ejf,Sucu:2026nkw}.The line element of the metric describing the Kalb-Ramond-ModMax spacetime is given by \cite{Sekhmani:2025jbl}
 
 \begin{equation}\label{1}
 	\mathrm{d}s^2= -f(r)\mathrm{d}t^2 + \frac{\mathrm{d}r^2}{f(r)} +r^2\left(\mathrm{d}\theta^2+\sin^2\theta \,{\mathrm{d}\phi^2}\right)\ ,
 \end{equation}
 where
\begin{equation}\label{m1}
f(r)= \frac{1}{1-l}-\frac{2M}{r}+\frac{\xi{Q^2}e^{-\gamma}}{(1-l)^2r^2} \ .
\end{equation}

 In the Eq. (\ref{m1}),  the parameters $M$ and $Q$ represent the mass and electric charge of the BH, respectively. {The discrete parameter $\xi$ determines whether the field is {phantom} $\xi=-1$ or canonical $\xi=1$.}
 The effect associated with the violation of Lorentz symmetry arises naturally due to the presence of the antisymmetric Kalb-Ramond stress, which is associated with the expected non-zero vacuum value and is represented by the small parameter $l$. On the other hand, we have that the parameter $\gamma$ is defined as positive and controls the degree of non-linearity of electromagnetism in such a way that when it tends to zero $\gamma\to{0}$ Maxwell's electromagnetism is recovered.
 
The validity regime of the metric tensor coordinates is defined by: the coordinates $t$ and $r$ associated with the line element Eq. (\ref{1}) and defined in Eq. (\ref{m1}) $\in$ to the interval $(-\infty,\infty)$, $\theta$ $\in$ $[0,\pi)$ and $\phi$ $\in$ $[0,2\pi)$. It is worth highlighting that, in the asymptotic regime, $f(r) \to 1-\frac{l}{l-1}$. This implies that this metric describes a spacetime with a topological charge, just like in the global Monopole spacetime of Barriola and Vilenkin \cite{GMVILENKIN}. Thus, depending on the value of $l$, it is a space with an angular deficit/surplus effect, which influences photon propagation, as we will discuss later.

 In a concise manner, we can define an expression to describe the position of the horizon structure, which is given below:

\begin{equation}\label{8}
	r_{H_{\pm}}=M(1-l)\Bigg[1\pm\sqrt{1-\frac{\xi{Q^2}e^{-\gamma}}{M^2(1-l)^3}}\Bigg],
\end{equation}
where the positive sign represents the position of the event horizon $r_{H_+}$ and the negative sign the Cauchy horizon $r_{H_-}$. By definition, for an event horizon to exist in the case of the BH with canonical field $\xi=1$, the argument inside the square root must be positive and definite, imposing the following restriction among the system parameters $ 0\leq{Q}\leq{M{e^{\gamma/2}}(1-l)^{3/2}}$. Note that we can rewrite the above constraint as $Q^2/M^2\leq(1-l)^3{e^\gamma}$ so that when the Lorentz violation parameters and the ModMax electrodynamics parameter cease to act $l=0=\gamma$, we then have the case of the extreme Reissner–Nordström BH $M=Q$ and the non-extreme $M>Q$ \cite{Duan:2023gng}.

In order for the horizons to be well defined, two basic requirements must be satisfied:  reality of the square root and positivity of the resulting radii. First, the reality condition demands
\begin{equation}
1-\frac{\xi Q^{2} e^{-\gamma}}{M^{2}(1-l)^{3}} \ge 0,
\end{equation}
which can be rewritten as
\begin{equation}
\xi Q^{2} e^{-\gamma} \le M^{2}(1-l)^{3}.
\end{equation}
Defining
\begin{equation}
\Delta \equiv 
1-\frac{\xi Q^{2} e^{-\gamma}}{M^{2}(1-l)^{3}},
\end{equation}
the condition $\Delta \ge 0$ ensures that the horizons are real. The limiting case $\Delta=0$ corresponds to an extremal configuration in which the two horizons coincide.

Second, the positivity of $r_{H_{\pm}}$ requires that the overall prefactor $M(1-l)$ be positive. Assuming a positive mass parameter $M>0$, this implies
\begin{equation}
1-l > 0 \qquad \Longleftrightarrow \qquad l < 1.
\end{equation}
Under this condition, and provided $\Delta \ge 0$, both radii are non–negative. In particular, for $0<\Delta<1$, one has two distinct horizons,
\begin{equation}
r_{H_{+}} > r_{H_{-}} > 0,
\end{equation}
whereas $\Delta=0$ yields the degenerate (extremal) case
\begin{equation}
r_{H_{+}} = r_{H_{-}} = M(1-l).
\end{equation}
If $\Delta<0$, the square root becomes imaginary and no real horizons exist. Therefore, the necessary and sufficient conditions for real, positive event horizons are
\begin{equation}
M>0, 
\qquad 
\ell<1,
\qquad
\xi Q^{2} e^{-\gamma} \le M^{2}(1-l)^{3}.
\end{equation}

\begin{figure}
    \centering
    \includegraphics[scale=0.55]{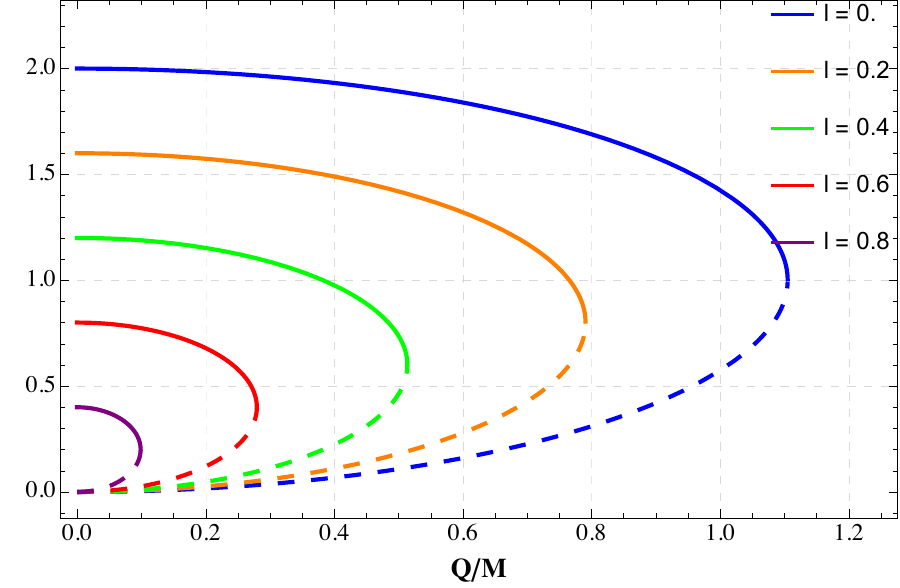}
        \includegraphics[scale=0.55]{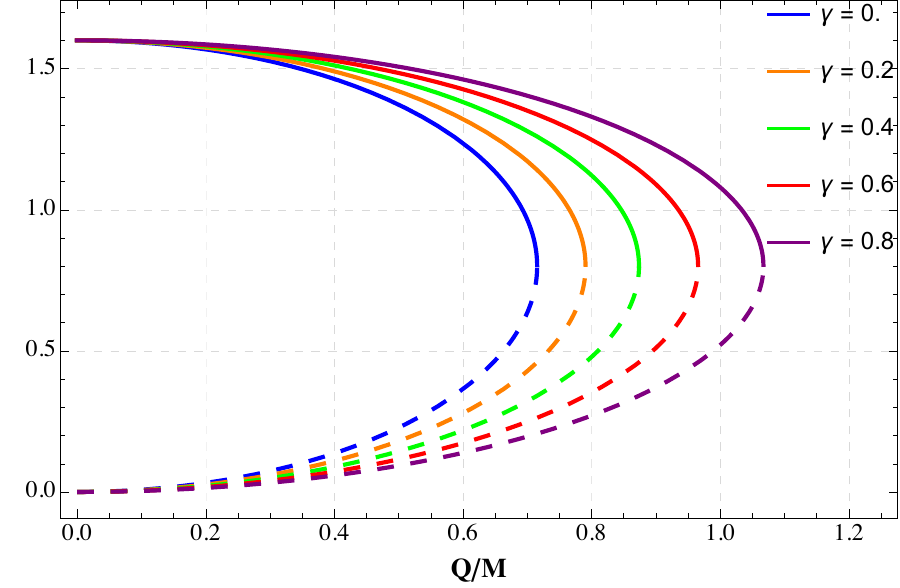}
        \includegraphics[scale=0.55]{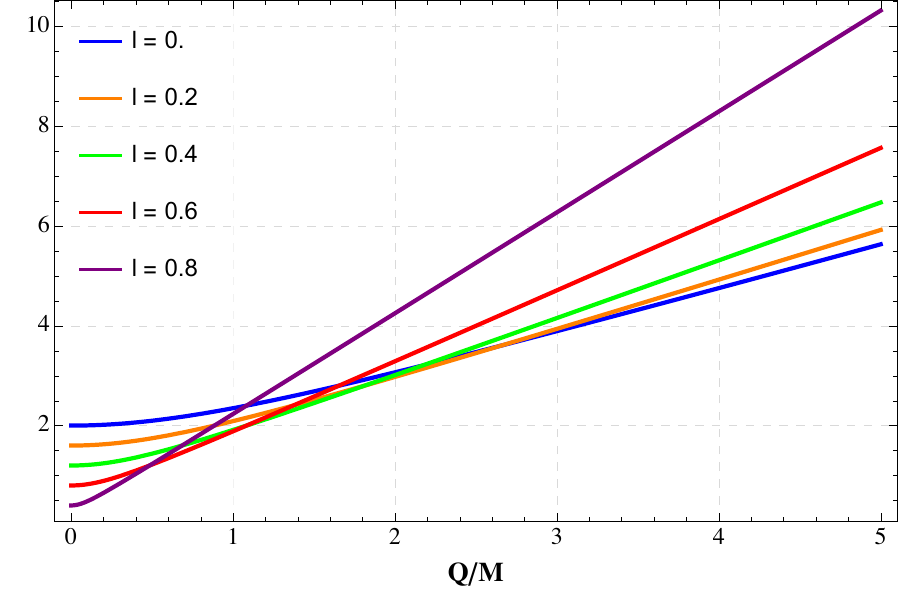}
        \includegraphics[scale=0.55]{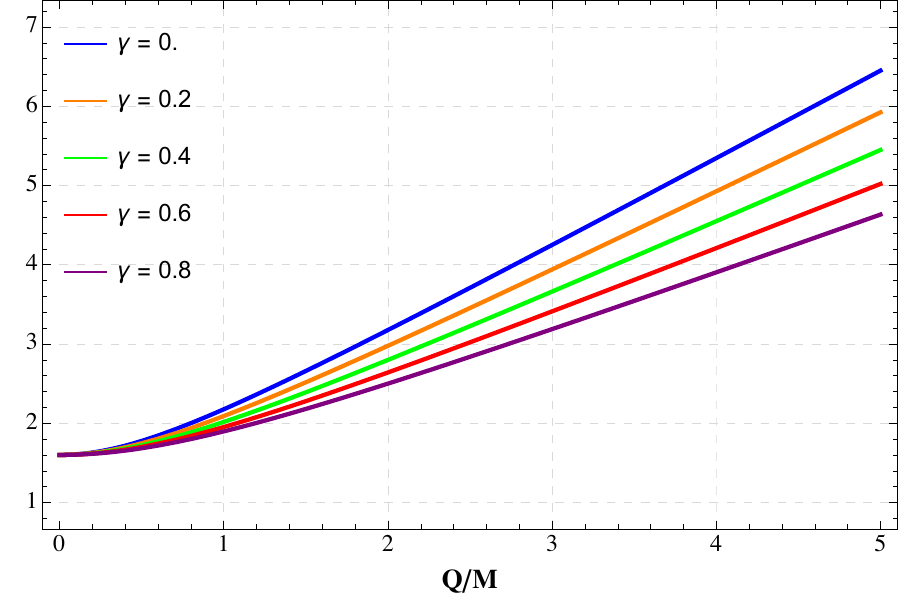}
    \caption{{Behavior of the event horizon radius and the Cauchy horizon radius as the black hole charge is varied for different values of the parameters. In the upper panels, we consider the canonical scalar field case, corresponding to $\xi=1$, while in the lower panels we consider the phantom case, $\xi=-1$. In the left panels, we fix $\gamma=0.2$ and vary the values of $l$, whereas in the right panels, we fix $l=0.2$ and vary the values of $\gamma$.}}
    \label{Horizon}
\end{figure}

{To complement this analysis, we investigate how the horizons behave when the parameters of the model are varied. In Fig.~(\ref{Horizon}), we change the model parameters while varying the black hole charge in order to examine the behavior of the horizons. We find that, for the case $\xi=1$, the behavior is similar to that of the Reissner--Nordström solution, where both an event horizon and a Cauchy horizon are present. As the charge increases and exceeds its extremal value, both horizons disappear, leaving a naked singularity. We also observe that increasing the parameter $l$ decreases the maximum charge allowed for the existence of horizons. Therefore, the extremal charge decreases as $l$ increases. On the other hand, when varying $\gamma$, we find that increasing $\gamma$ raises the limiting value of the electric charge, meaning that the extremal charge is enhanced by the presence of the parameter $\gamma$.}

{For the phantom field case, $\xi=-1$, we observe that there is no Cauchy horizon and only an event horizon is present. Moreover, the existence of the horizon is no longer limited by the value of the black-hole charge. The radius of the event horizon increases as the charge is increased. The solid curves correspond to the radius of the event horizon, whereas the dashed curves represent the radius of the Cauchy horizon.}

Now, to study the deflection of light let us consider the
	geodesics of particles moving in the Kalb-Ramond-ModMax spacetime.
	Thus, we define a smooth curve in this  spacetime, Eq. \eqref{1}, that has length $S$ which is given by
	\begin{equation}\label{2}
		S= \int \sqrt{\left(g_{\mu\nu}\frac{\mathrm{d}x^\mu}{\mathrm{d}\lambda}\frac{\mathrm{d}x^\nu}{\mathrm{d}\lambda}\right)}\mathrm{d}\lambda,
	\end{equation}
    where $\lambda$ is an affine parameter that can represent the observer's proper time. Taking $S$ as the affine parameter itself, we can show that the curve that minimizes Eq. (\ref{2}) is the same one that minimizes
	\begin{equation}\label{3}
		\int \left(g_{\mu\nu}\frac{\mathrm{d}x^\mu}{\mathrm{d}\lambda}\frac{\mathrm{d}x^\nu}{\mathrm{d}\lambda}\right)\mathrm{d}\lambda=\int\mathcal{L}\,{\mathrm{d}\lambda}.
	\end{equation}
	
	Therefore, considering the analysis in the equatorial plane, $\theta=\frac{\pi}{2}$, the Lagrangian becomes:
	\begin{equation}\label{4}
		\mathcal{L}= -f(r)\left(\frac{\mathrm{d}t}{\mathrm{d}\lambda}\right)^2 + \frac{1}{f(r)}\left(\frac{\mathrm{d}r}{\mathrm{d}\lambda}\right)^2 + r^2\left(\frac{\mathrm{d}\phi}{\mathrm{d}\lambda}\right)^2.
	\end{equation}
	
	Applying the Euler-Lagrange equations to the above expression, we define the quantities conserved at time $t$ and at $\phi$. Therefore, we have to
	\begin{equation}\label{5}
		L= r^2\frac{\mathrm{d}\phi}{\mathrm{d}\lambda},  \qquad \qquad E = f(r)\frac{\mathrm{d}t}{\mathrm{d}\lambda}.
	\end{equation}
	
	Thus, substituting the conserved quantities, Eq. \eqref{5}, in Eq. \eqref{4} and considering only null geodesics, $\mathcal{L}=0$, it leads (\ref{4}) to
	\begin{equation}\label{6}
		\left(\frac{\mathrm{d}r}{\mathrm{d}\lambda}\right)^2= \left[E^2- \frac{L^2{f(r)}}{r^2}\right].
	\end{equation}

	The expression above may be interpreted by analogy with the dynamics of a classical particle of unit mass, with energy $\mathcal{E}$, subject to an effective potential $V_{\text{eff}}$, given by
	\begin{equation}\label{6A}
		\mathcal{E}=E^2,\qquad\text{and}\qquad V_{\text{eff}}=\frac{L^2f(r)}{r^2} \ .
	\end{equation}
	
	In terms of Eq. (\ref{m1}), the effective potential becomes
	\begin{equation}\label{m2}
		 V_{\text{eff}}=  \frac{L^2}{r^2} \left(	\frac{1}{1-l}-\frac{2M}{r}+\frac{\xi{Q^2}e^{-\gamma}}{(1-l)^2r^2} \right).
	\end{equation}
	
	Radial motion is permitted only when $\mathcal{E}>V_{\text{eff}}$. In this context, we examine the situation in which the photon begins its trajectory in an asymptotically flat region and approaches a distance 
	$r_0$ from the center of BH, reaching the so-called turning point, located outside the event horizon. As expected, once the photon reaches this limiting region, the gravitational field causes it to return toward another asymptotically flat region. At the turning point, one has, $V_{\text{eff}}(r_0)=E^2$, which, from Eq. (\ref{6A}), leads to
	\begin{equation}\label{6B}
		\frac{1}{\beta^2}=\frac{f(r_0)}{r^2_0} \ ,
	\end{equation}
	where $\beta=\frac{L}{E}$ is the  impact parameter of the light ray. At the photon sphere radius, $r_m$, we have
	\begin{equation}\label{6C}
		\frac{\mathrm{d}V_{\text{eff}}}{\mathrm{d}r}\Bigg|_{r_m}=0 \ .
	\end{equation}
    
	The critical impact parameter, $\beta_c$, is defined by the condition 
	$\beta_c=\beta(r_m)$. Light rays with $\beta<\beta_c$ are completely absorbed, whereas those for which $\beta=\beta_c$ remain trapped on the photon sphere; in contrast, rays with $\beta>\beta_c$ are deflected and ultimately scattered. In the scattering regime, the deflection angle diverges in the limit $r_0 \to{r_m}$, which characterizes the so-called strong-field regime. In the following sections, we derive the approximation for the deflection angle in these two regimes, for this new class of electrically charged BH within the Kalb-Ramond theory.

Using Eq. (\ref{6C}) and Eq. (\ref{m2}) , we can identify the position of the radius of the photon sphere, which is given in the expression below and has an inner $r_{m1}$ and an outer radius $r_{m2}$
\begin{equation}\label{9}
r_{m1}=\frac{3M(1-l)}{2}\Bigg[1-\sqrt{1-\frac{8D}{9}}\Bigg], \qquad r_{m2}=\frac{3M(1-l)}{2}\Bigg[1+\sqrt{1-\frac{8D}{9}}\Bigg],
\end{equation} where the parameter above is defined as $D=\frac{\xi{Q^2}{e^{-\gamma}}}{M^2(1-l)^3}$. Note that in the limit where the electric charge tends to zero $Q\to{0}$ the radius of the photon sphere is recovered for the Schwarzschild BH, being deformed by the Lorentz symmetry violation parameter $r_{m2}=3M(1-l)$  \cite{Junior:2024vdk} .


\section{\label{sec4}Lensing phenomena}

\subsection{Expansion for light  deflection in the weak-field limit}\label{sec31}

Considering light deflection in this scenario, with the line element and metric function defined by Eqs. (\ref{1}) and (\ref{m1}) respectively, the expression for the impact parameter is obtained from Eq. (\ref{6B}) as follows:
\begin{equation}\label{10}
    \frac{1}{\beta^2(r_0)}= \frac{1}{r^2_0}\Bigg[\frac{1}{(1-l)}-\frac{2M}{r_0}+\frac{(1-l)DM^2}{r^2_0}\Bigg].
\end{equation}
In terms of Eq. (\ref{5}), the radial equation Eq. (\ref{6}) becomes
\begin{equation}\label{12}
	\Big(\frac{\mathrm{d}r}{\mathrm{d}\phi}\Big)^2= r^4\left[\frac{E^2}{L^2}-\frac{f(r)}{r^2} \right] .
\end{equation}

Considering the symmetry that the distances before and after the inflection point are equal, we see that the contributions to the angular deviation are the same; therefore, we multiply the expression for the angular deviation by the factor 2;
\begin{equation}\label{13}
    \Delta\phi= 2\,\int^{\infty}_{r_0} \frac{dr}{r^2}\Bigg[\frac{1}{\beta^2}-\frac{f(r)}{r^2}\Bigg]^{-1/2},
\end{equation} in developing this work, for the sake of symmetry, we only adopted the positive sign from the expression above. We now introduce a coordinate transformation in Eq. (\ref{13}), where $u=\frac{1}{r}$. Note that, under this transformation, the integration limits become $u\to{0}$ as $r\to\infty$, and $u\to{u_0}=\frac{1}{r_0}$ as $r\to{r_0}$. Consequently, the equation for the angular deviation in the new coordinates takes the form
\begin{equation}\label{14}
    \Delta\phi= \int^{u_0}_0{\mathrm{d}u}\left[\frac{1}{\beta^2(r_0)}-\frac{u^2}{(1-l)}+2Mu^3-(1-l)DM^2u^4\right]^{-1/2}.
\end{equation}

Since our interest is in performing the angular deviation expansion of light in the weak-field regime, we must return to the expression for the impact parameter defined in Eq. (\ref{10}) and then apply the same coordinate transformation introduced above, leading to the expression $\frac{1}{\beta^2}=u^2_0\Bigg[\frac{1}{(1-l)}-2Mu_0 + (1-l)DM^2u^2_0\Bigg]$. Using this new form of the impact parameter and assuming that the photon travels away from the BH, we can adopt the approximation that both mass and electric charge are small parameters $M<<1$ and $Q<<1$. Consequently, expanding the orbital equation Eq. (\ref{14}), up to the second order in $M$ and charge $Q$, we find that the deflection of light is given by $\delta\phi = \Delta\phi - \pi$\footnote{{Here, $\Delta\phi$ denotes the total change in the angular coordinate along the photon trajectory from infinity to infinity. Since a straight-line path in flat spacetime corresponds to $\Delta\phi=\pi$, the deflection angle is given by $\delta\phi=\Delta\phi-\pi$.}}  and for the purposes of graphical analysis we will discard cross terms higher than the second order:
\begin{eqnarray}\label{15}
    \delta\phi_{KR-ModMax}&\approx& \pi\Big(-1+\sqrt{1-l}\Big) + \frac{4M(1-l)^{2}}{\beta} + \frac{{M^2}(15\pi-16)(1-l)^{7/2}}{4\beta^2} \nonumber \\
    &-& \frac{\xi{Q^2}e^{-\gamma}(1-l)^{1/2}}{32\beta^2}\Bigg[24\pi+ \frac{M(448-48\pi)(1-l)^{3/2}}{\beta}+\frac{25M^2(33\pi-64)(1-l)^3}{\beta^2}\Bigg]. \nonumber \\
\end{eqnarray}

In the expression above, we have the calculation of the light deviation for an electrically charged BH through nonlinear electrodynamics in the Kalb-Ramond theory, where the Lorentz symmetry violation effect is added through the non-zero expected vacuum value. Note that in the limit where the electric charge of the BH tends to zero $Q\to{0}$, we obtain the angular deviation for the Schwarzschild BH being deformed by the Lorentz symmetry violation parameter, which also comes from Kalb-Ramond gravity and can be seen in the expression below:
\begin{equation}\label{16}
    \delta\phi_{Schwarzschild-like}\approx \pi\Big(-1+\sqrt{1-l}\Big) + \frac{4M(1-l)^{2}}{\beta} + \frac{{M^2}(15\pi-16)(1-l)^{7/2}}{4\beta^2}.
\end{equation}

Regarding the angular deviation in the Schwarzschild BH limit being modified by the Lorentz symmetry violation effect Eq. (\ref{15}), the factor accompanying the second-order term in the mass $\mathcal{O}(M^2)$ is consistent with Refs. \cite{Gao:2024ejs,Keeton:2005jd}. On the other hand, when we consider the limit in which both the Lorentz violation effect and electrodynamics are off $l=\gamma=0$ and $\xi=1$, we recover the factor that multiplies the quadratic term in the electric charge for the Reissner-Nordström BH $\mathcal{O}(Q^2)$, being in accordance with Refs.\cite{Sasaki:2023rdf,Kumaran:2022soh}.

The expression for the simplified scenario above must be consistent with the light deviation in the field regime for a BH similar to that of Schwarzschild and analyzed in Ref. \cite{Junior:2024vdk}. On the other hand, by also turning off the Lorentz symmetry violation effects $l\to{0}$, we recover the light deviation for the Schwarzschild BH, and this result can be compared with several works obtained in the literature, and its reduced expression is given below
\begin{equation}\label{17}
    \delta\phi_{Schwarzschild}\approx  \frac{4M}{\beta} + \frac{{M^2}(15\pi-16)}{4\beta^2}.
\end{equation}

\begin{figure}[htb!]
\centering
	{\includegraphics[width=0.7\textwidth]{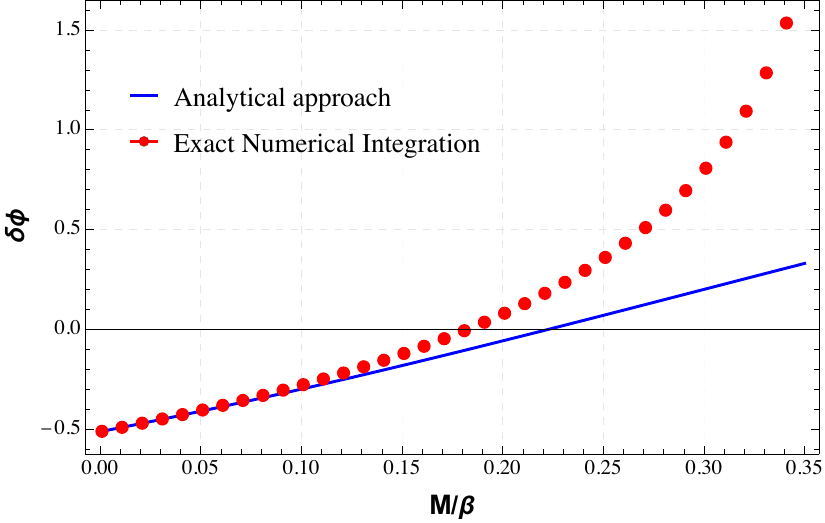}}
    \caption{{Comparison between the approximate analytical result given by Eq.~\eqref{15} and the numerical result obtained from Eq.~\eqref{13} for the deflection angle. For this comparison, we adopted $\xi=1$, $\gamma=0.2$, $l=0.2$, and $Q/M=0.5$.}}
\label{Numerico_analitico}
\end{figure}

{In order to assess the validity of the approximation given in Eq.~\eqref{15}, we compare it graphically with the numerical integration of Eq.~\eqref{13}. This comparison is shown in Fig.~(\ref{Numerico_analitico}). We find that the approximation provides an excellent agreement for small values of $M/\beta$. However, as this ratio increases, the analytical expression begins to exhibit significant discrepancies with respect to the numerical result.}

\begin{figure}[htb!]
\centering
	{\includegraphics[width=0.75\textwidth]{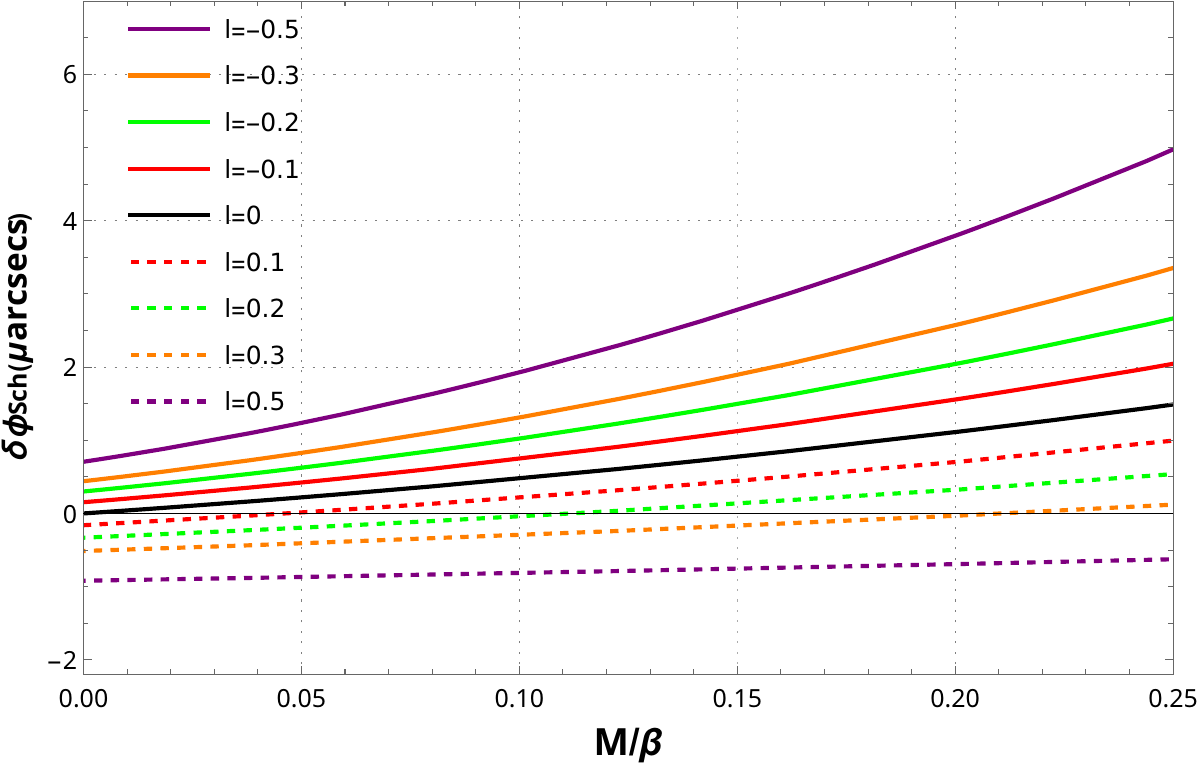}}
    \caption{Light deflection in the weak-field regime.}
\label{FRACO1}
\end{figure}

Figure (\ref{FRACO1}) provides a simplified illustration of weak-field light deflection for a Schwarzschild BH under Lorentz symmetry violation. This phenomenon, arising from Kalb-Ramond gravity, is governed by a parameter linked to a non-zero vacuum expectation value. Our results in Eq. (\ref{16}) remain consistent with the optical analysis previously established in Ref. \cite{Junior:2024vdk}. {In Fig. (\ref{FRACO1}), the solid black curve corresponds to standard Schwarzschild spacetime, while the dashed lines represent the influence of this vacuum-driven term. We observe that as this parameter increases positively, the angular deviation decreases, suggesting a potential attenuation of spacetime curvature or perhaps even representing a repulsive gravity regime. On the other hand, for small negative values, the angular deviation becomes positive, causing the metric to retain its signature Eq. (\ref{m1}). This behavior can be observed through the colored solid curves in the same figure. These results are consistent with Refs. \cite{Duan:2023gng, Junior:2024vdk, Cordeiro:2025cfo}.}

\begin{figure}
    \centering
    \includegraphics[scale=0.41]{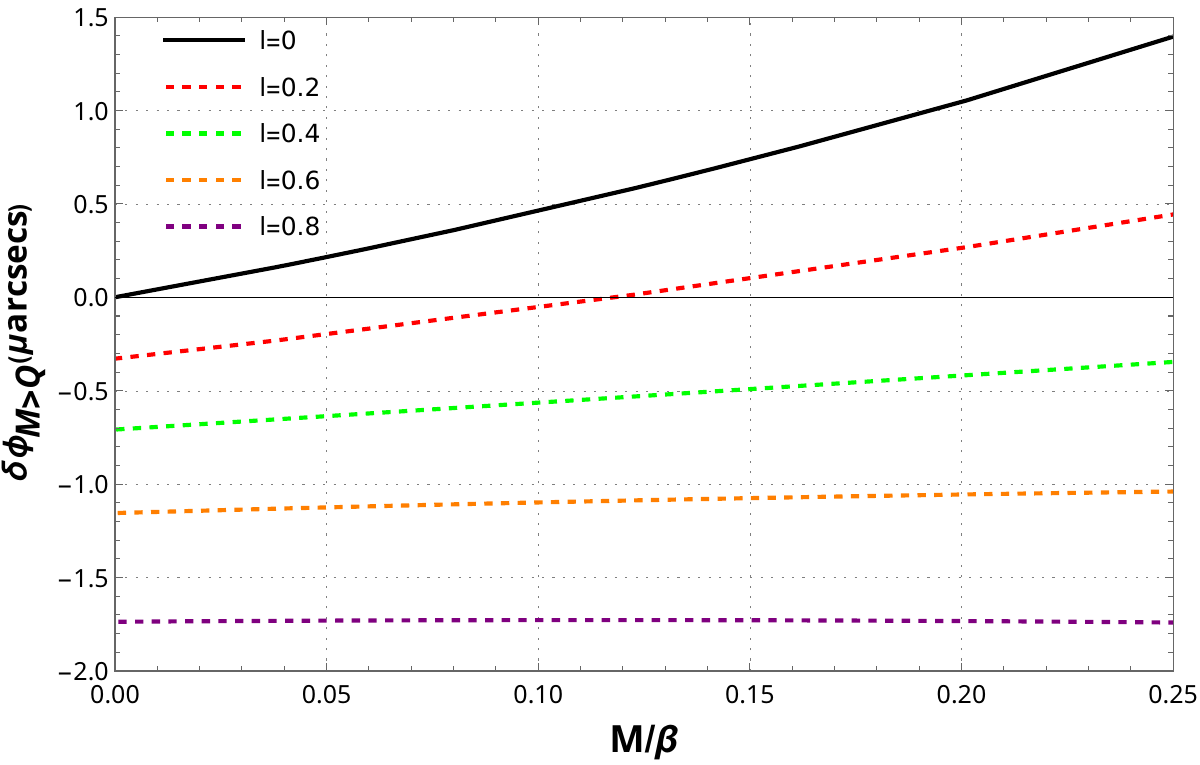}
    \includegraphics[scale=0.41]{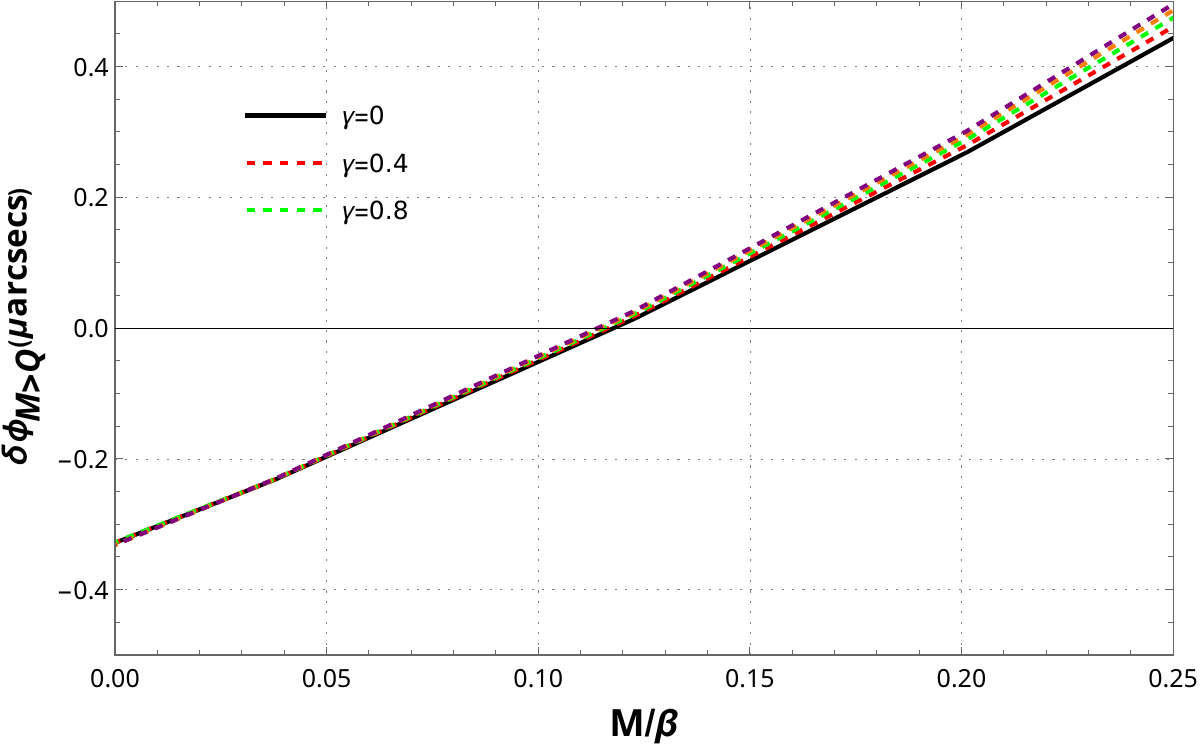}
    \caption{In the figures above, we fix the parameter $\xi=1$ and the mass $M=1.2Q$. In the panel on the left, we consider $\gamma=0$ and then vary $l$, while in the panel on the right we fix $l=0.2$ and then vary $\gamma$.}
    \label{FRACO2}
\end{figure}

\begin{figure}
    \centering
      \includegraphics[scale=0.41]{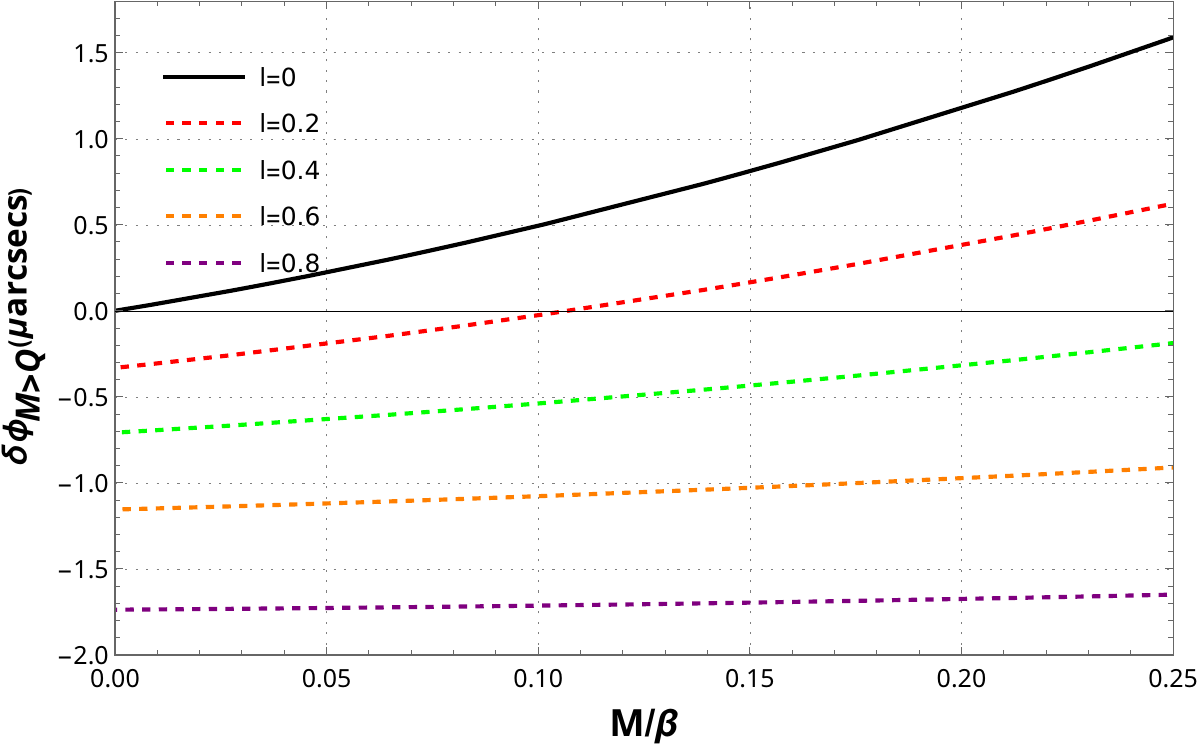}
        \includegraphics[scale=0.41]{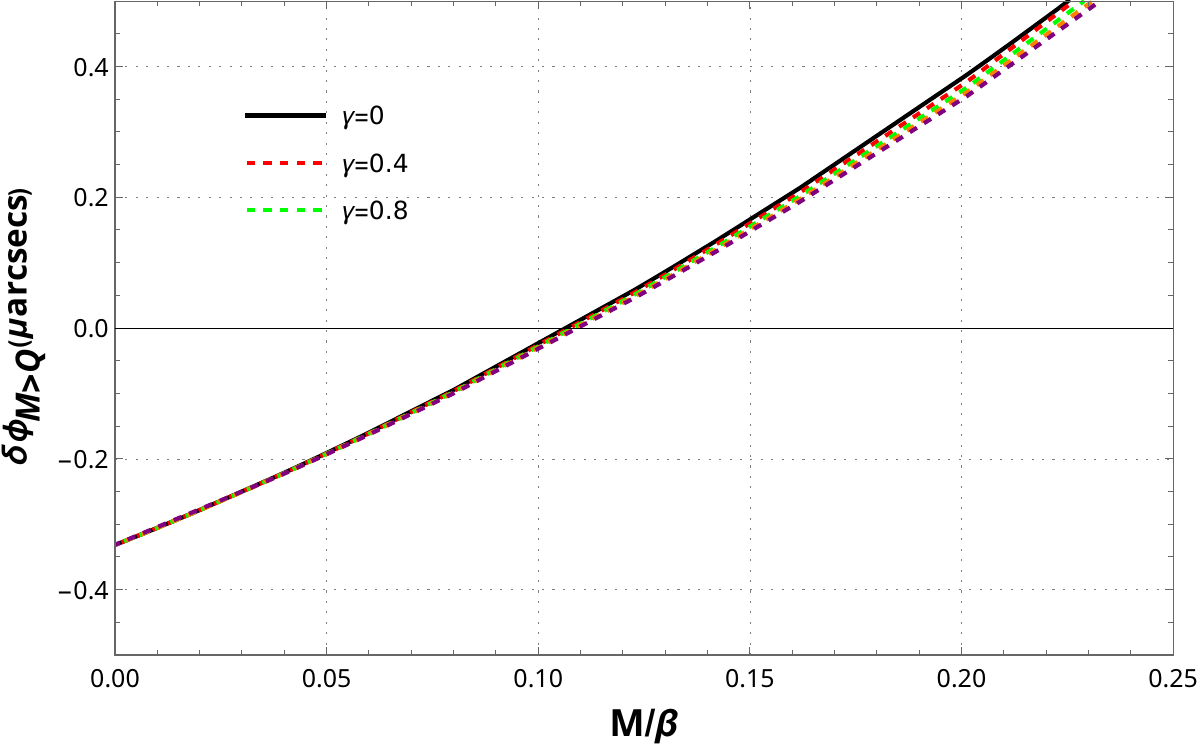}
    \caption{In the figures above, we fix the parameter $\xi=-1$ and the mass $M=1.2Q$. In the panel on the left, we consider $\gamma=0$ and then vary $l$, while in the panel on the right we fix $l=0.2$ and then vary $\gamma$.}
    \label{FRACO3}
\end{figure}

In Figs. (\ref{FRACO2}) and (\ref{FRACO3}), we have, respectively, the representation of the non-extreme BH, where its mass has been fixed proportionally to the electric charge as $M=1.2Q$, with the canonical field $\xi=1$ and the phantom field $\xi=-1$. Thus, we can compare the panels on the left of Figs. (\ref{FRACO2}) and (\ref{FRACO3}), where we varied the Lorentz symmetry violation parameter and kept $\gamma=0$. In both cases, the solid black curve represents the Schwarzschild BH and the dashed curves represent the deviation due to the influence of the Lorentz symmetry violation. In the upper left panel, we consider the canonical case and, in the lower left panel, the phantom case. {Observe in both left panels that there is not much variation between the curves, and therefore it is not possible to have an explicit perception of behavior.} 

On the other hand, in the panels on the right, in Figs. (\ref{FRACO2}) and (\ref{FRACO3}), we fixed the Lorentz symmetry violation parameter and then varied the parameter $\gamma$, which controls the nonlinearity of the electrodynamics. Thus, the solid black curve represents the angular deviation for $l=0.2$, and the dashed curves show how this behavior occurs as we increase the nonlinearity in the model. Note that the behavior of the canonical field for the phantom case changes only due to a matter of symmetry with respect to the solid black reference curve. 

\begin{figure}
    \centering
    \includegraphics[scale=0.41]{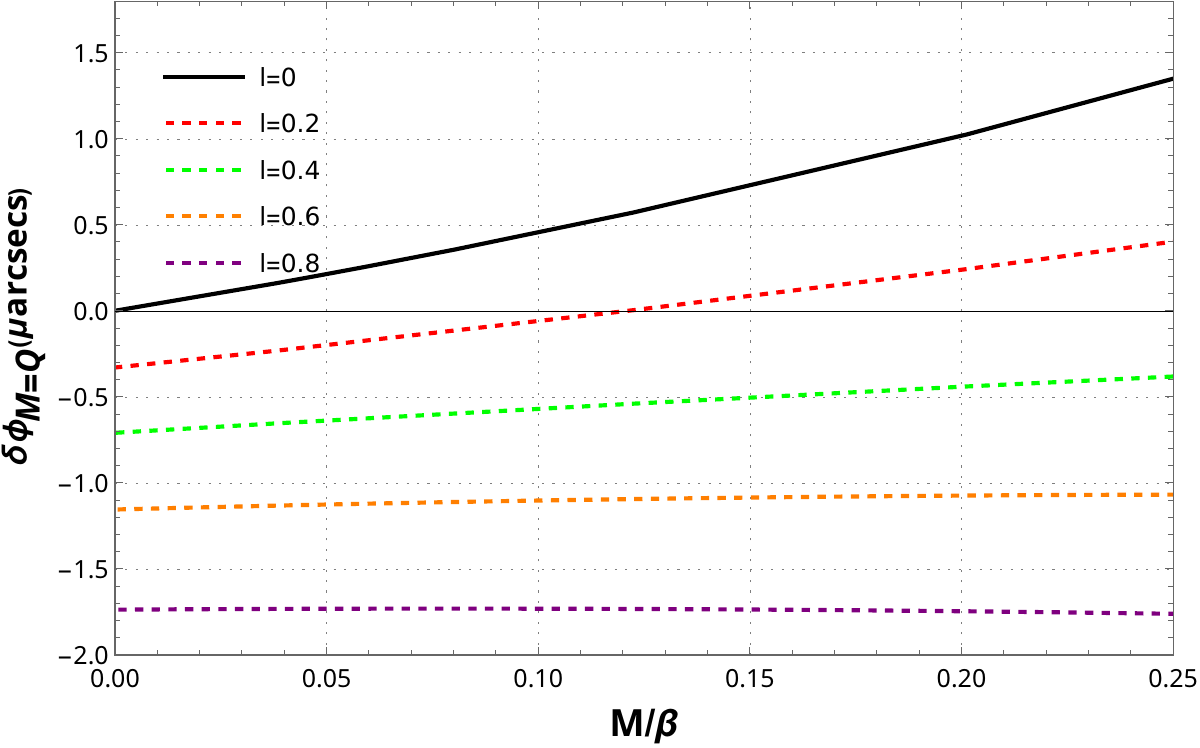}
    \includegraphics[scale=0.41]{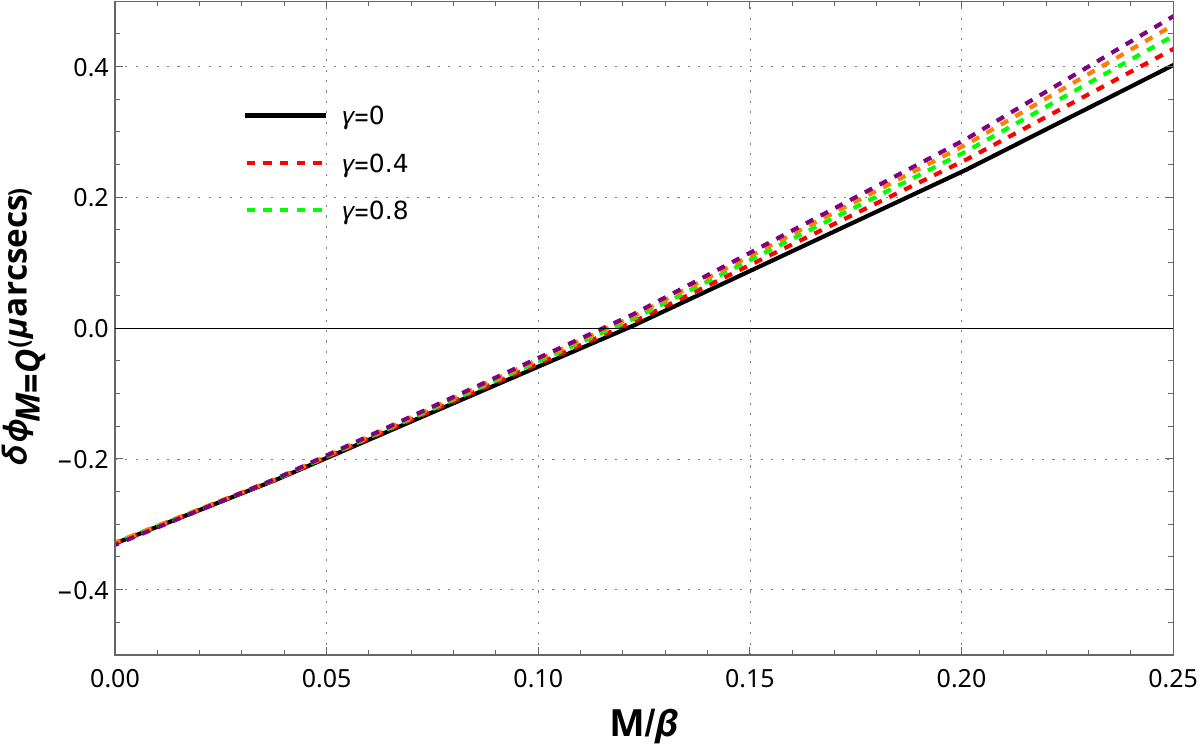}
    \caption{In the figures above, we fix the parameter $\xi=1$ and the mass $M=Q$. In the panel on the left, we consider $\gamma=0$ and then vary $l$, while in the panel on the right we fix $l=0.2$ and then vary $\gamma$.}
    \label{FRACO4}
\end{figure}

\begin{figure}
    \centering
      \includegraphics[scale=0.41]{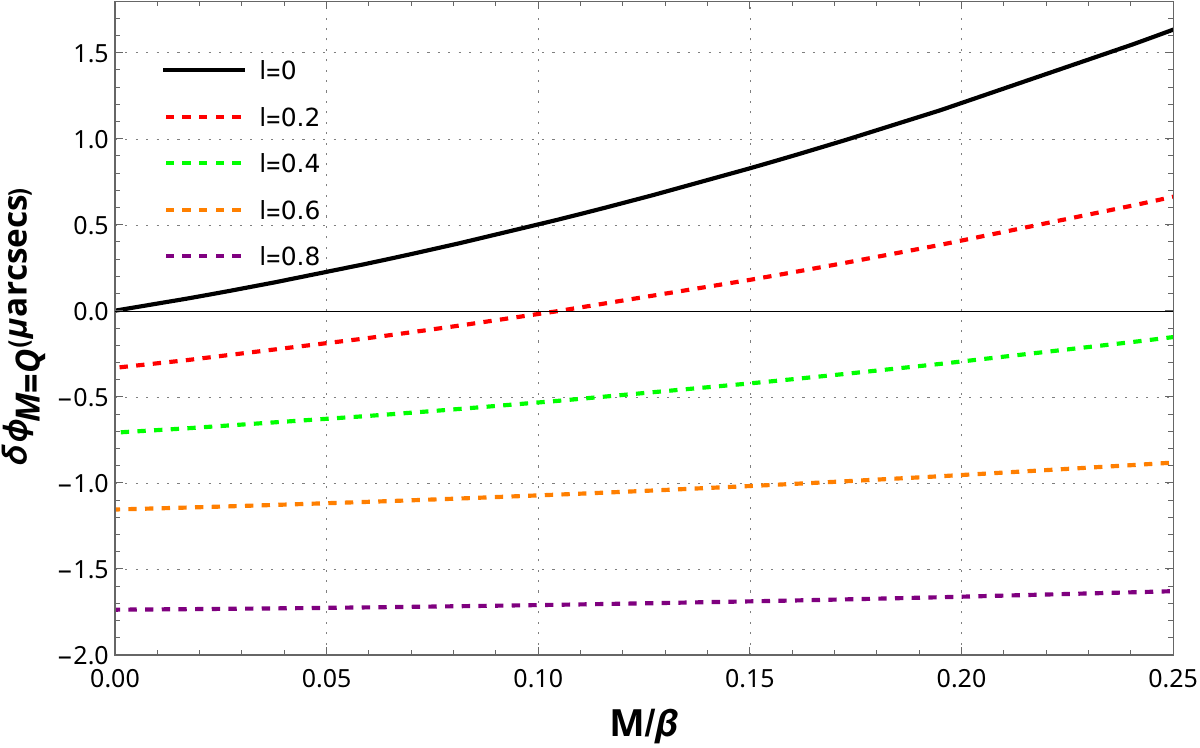}
        \includegraphics[scale=0.41]{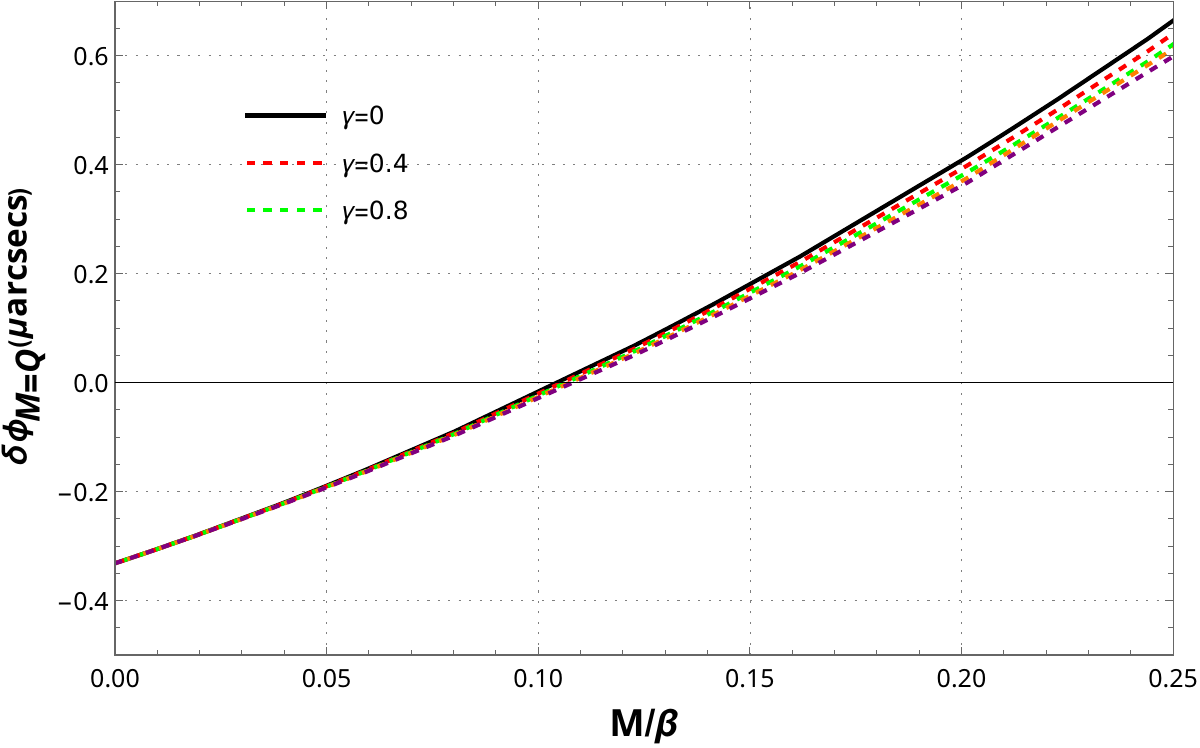}
    \caption{In the figures above, we fix the parameter $\xi=-1$ and the mass $M=Q$. In the panel on the left, we consider $\gamma=0$ and then vary $l$, while in the panel on the right we fix $l=0.2$ and then vary $\gamma$.}
    \label{FRACO5}
\end{figure}

\begin{figure}
    \centering
    \includegraphics[scale=0.41]{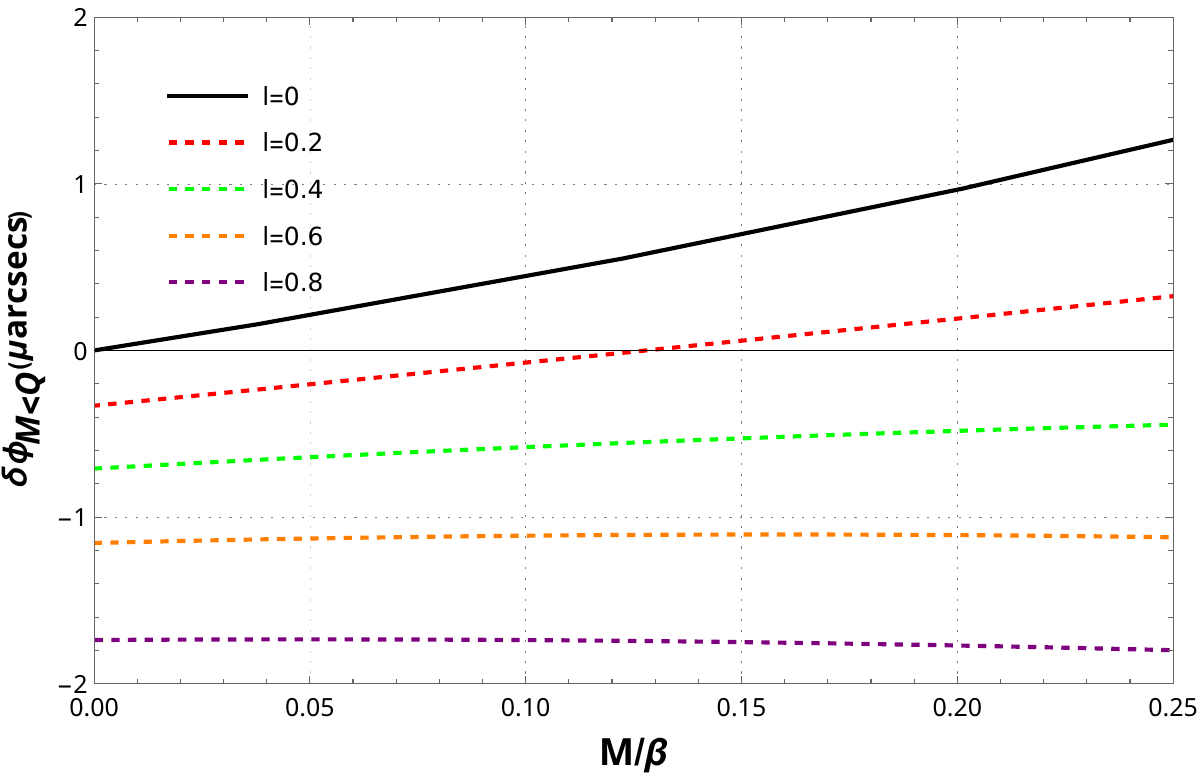}
    \includegraphics[scale=0.41]{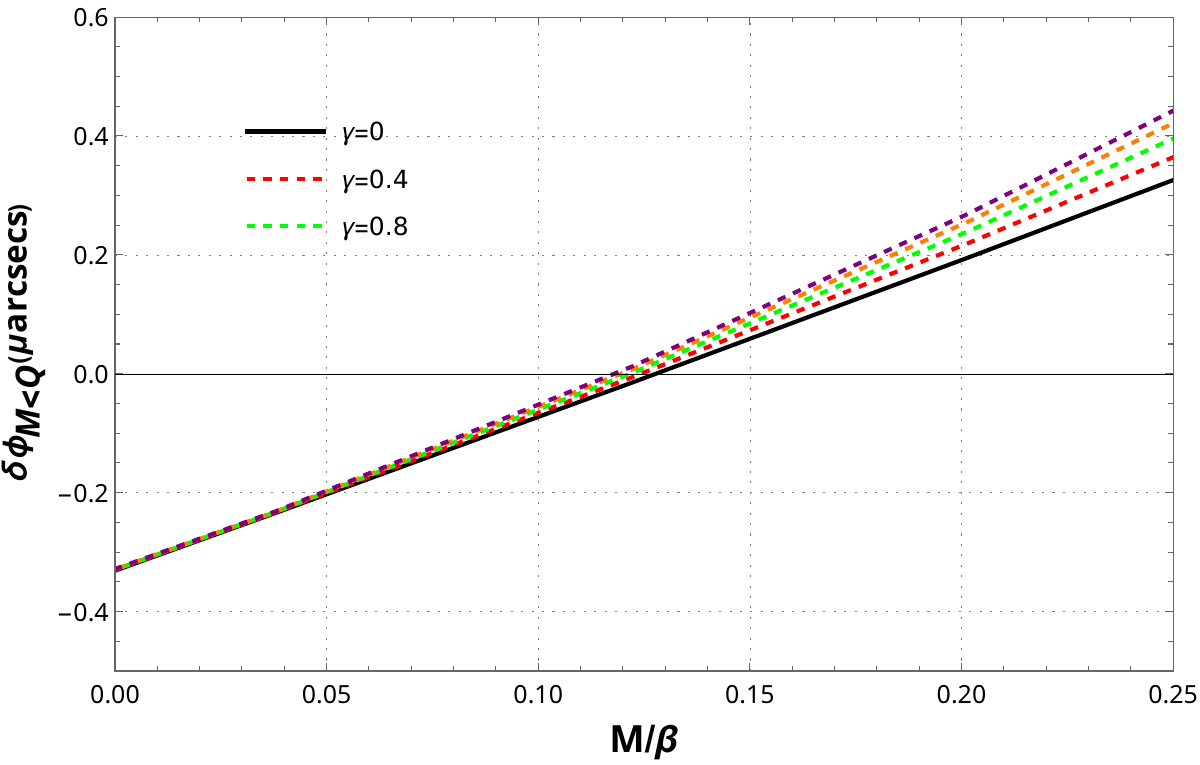}
    \caption{In the figures above, we fix the parameter $\xi=1$ and the mass $M=0.8Q$. In the panel on the left, we consider $\gamma=0$ and then vary $l$, while in the panel on the right we fix $l=0.2$ and then vary $\gamma$.}
    \label{FRACO6}
\end{figure}

\begin{figure}
    \centering
      \includegraphics[scale=0.41]{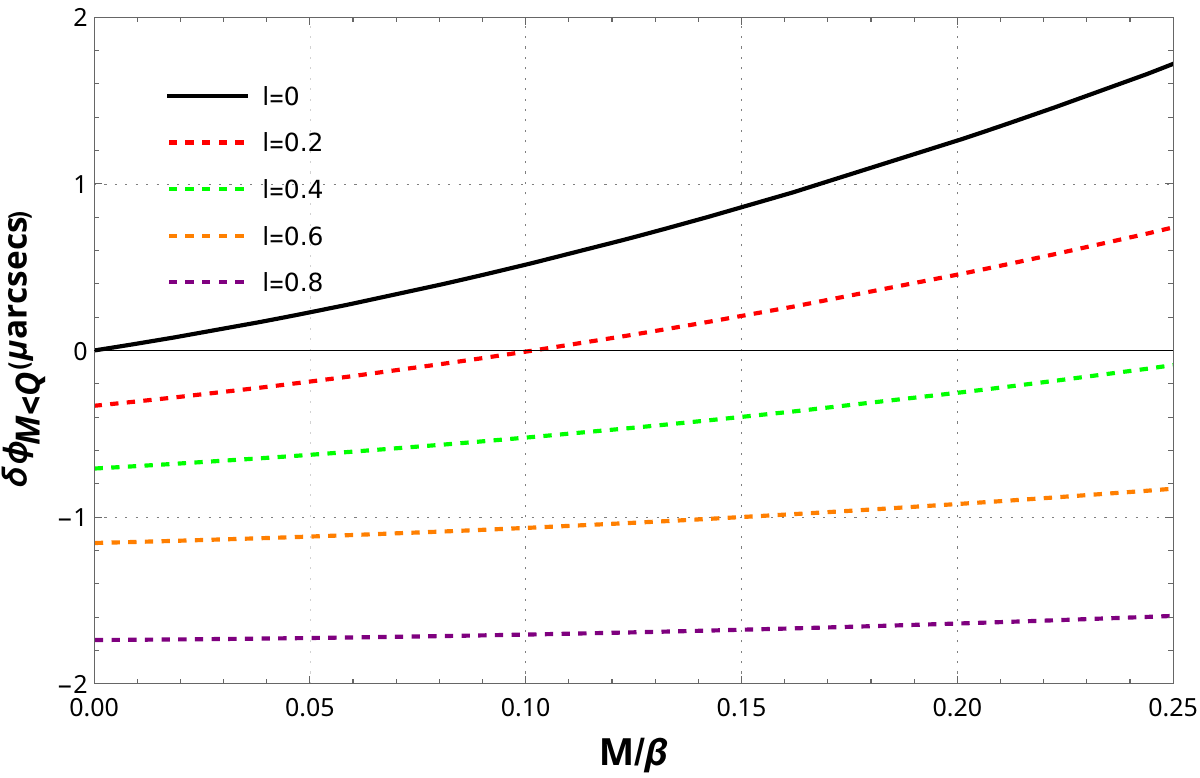}
        \includegraphics[scale=0.41]{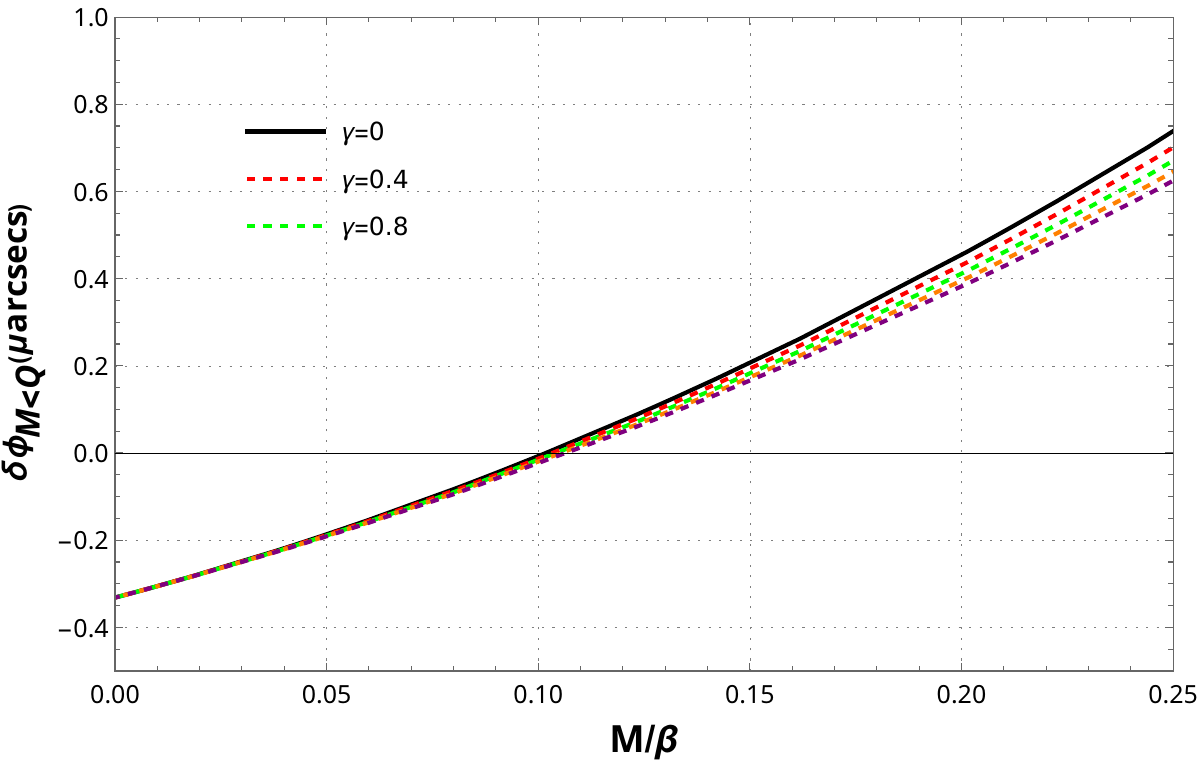}
    \caption{In the figures above, we fix the parameter $\xi=-1$ and the mass $M=0.8Q$. In the panel on the left, we consider $\gamma=0$ and then vary $l$, while in the panel on the right we fix $l=0.2$ and then vary $\gamma$.}
    \label{FRACO7}
\end{figure}

{Similar to the analysis performed for the non-extreme BH case, in the panels of Figs. (\ref{FRACO4}--\ref{FRACO7}) we perform the same procedure, now considering the configurations for the case $M=Q$ and naked singularity $Q>M$, for both the canonical and phantom cases. No significant change in behavior was observed, establishing a reasonably standardized pattern for angular deviation in these investigated scenarios.}

\begin{figure}
    \centering
      \includegraphics[scale=0.54]{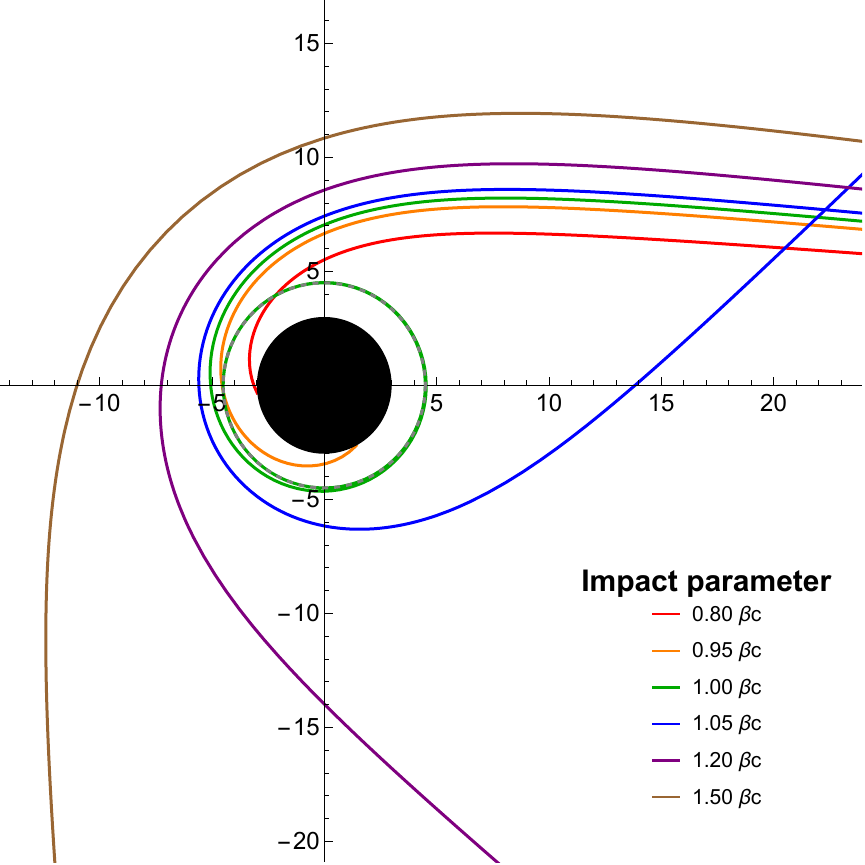}
        \includegraphics[scale=0.54]{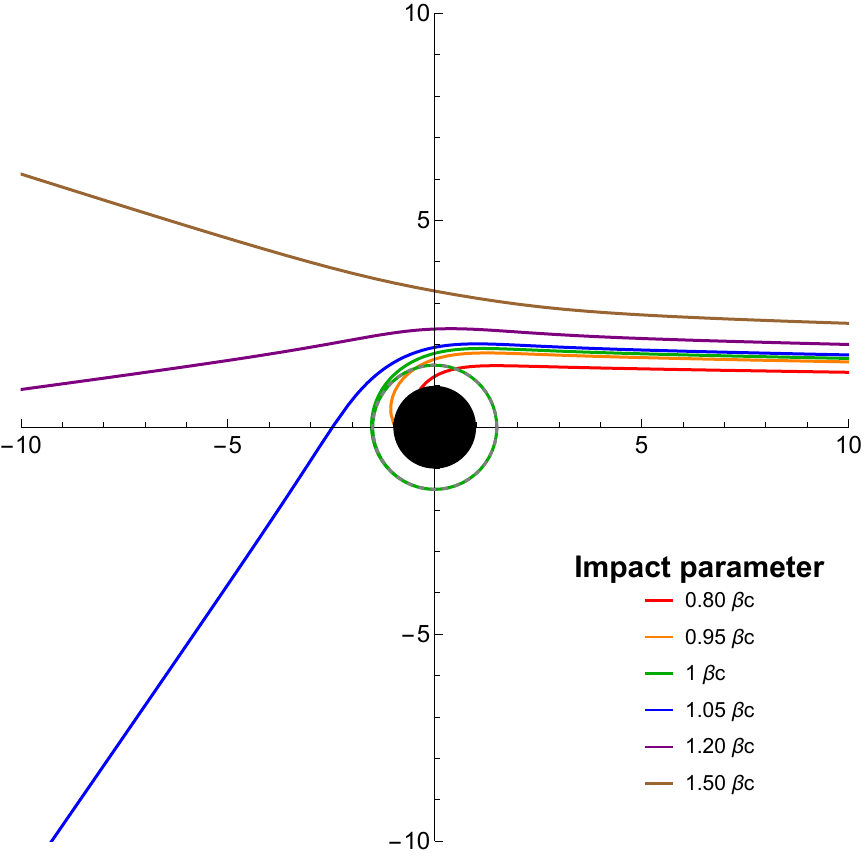
        }
    \caption{{Null geodesics for $Q=0$, $\xi=1$, and different values of the impact parameter $b$. The left panel shows the case $l=-0.5$, while the right panel shows the case $l=0.5$. The dark region in the figure represents the area enclosed by the black hole event horizon. If the impact parameter is smaller than the critical impact parameter, the particles are absorbed by the black hole. If the impact parameter is larger than the critical value, the particles are scattered. In the limiting case where the impact parameter is equal to the critical one, the particles undergo an unstable circular orbit.}}
    \label{Geodesics}
\end{figure}

{Both Figs.~(\ref{FRACO1}--\ref{FRACO7}) and Eqs.~\eqref{15} and \eqref{16} clearly show that the deflection angle becomes negative for sufficiently large values of the impact parameter when $l>0$. This is not a mathematical artifact of the weak-field expansions. In Fig.~\ref{Geodesics}, we analyze the behavior of massless particles for the case $Q=0$ for simplicity. We find that, for $l<0$, the deflection exhibits the usual behavior expected from an attractive gravitational field. However, when $l>0$ and the impact parameter is increased, the particles experience a deflection in the opposite direction to that normally expected, as if the gravitational field of the black hole were repelling them rather than attracting them.}

{As we discussed previously, the spacetime under consideration behaves asymptotically as the Barriola--Vilenkin solution, which can be written asymptotically as
\begin{equation}
    f(r\to \infty) \approx 1-\alpha^2,
\end{equation}
where $\alpha$ is a parameter related to the symmetry breaking. We may attempt to compare the weak-field limit of the light deflection angle in the Barriola--Vilenkin model with that obtained in our case. The expression for the weak-field deflection angle is given by
\begin{equation}
\delta\phi_{Schwarzschild-like}\approx \pi\Bigg(-1+\frac{1}{\sqrt{1-\alpha^2}}\Bigg) + \frac{4M}{\beta(1-\alpha^2)^{2}}+O\left(\frac{1}{\beta^2}\right).
\end{equation}
From this, we observe that there is an equivalence between the dominant terms in the weak-field regime through the substitution $\alpha^{2}\rightarrow l/(l-1)$. In the Barriola--Vilenkin case, the deflection angle never becomes negative since $\alpha$ is a real parameter. In our model, this situation is equivalent to having negative values of $l$. The negative values of the deflection angle in our case arise in the interval $0<l<1$, which would correspond to having an imaginary value of $\alpha$ in the Barriola--Vilenkin solution.}

\subsection{Deflection of light in the strong-field limit}\label{sec32}

In this section, our interest focuses on calculating the gravitational deflection of light in the strong-field regime, and for this, we will use as a basis the mechanism originally proposed by Bozza \cite{B8} and later improved by Tsukamoto \cite{B9}. In this context, we begin by transforming the variables $z=1-\frac{r_0}{r}$ into the orbit equation, Eq. (\ref{13}), from which it can be expressed again as:

\begin{equation}\label{18}
\Delta\phi(r_0)=\pm \int^1_0 \frac{2r_0{\mathrm{d}z}}{\sqrt{G(z,r_0)}},
\end{equation} where
\begin{equation}\label{19}
G(z,r_0)= \frac{r^4_0}{\beta^2}-\Bigg(\frac{r^2_0}{1-l}\Bigg)(1-z)^2+2Mr_0{(1-z)^3}-(1-l)DM^2(1-z)^4.
\end{equation}

After considering the series expansion of the function $G(z,r_0)$ at the point $z=0$, which corresponds to the limit $r\to {r_0}$, and considering $\beta=\beta(r_0)$ given by Eq. (\ref{6B}), we obtain that: 
\begin{equation}\label{20}
G(z,r_0) \simeq \Lambda_1(r_0)z + \Lambda_2(r_0)z^2,
\end{equation} where the expansion parameters $\Lambda_1$ and $\Lambda_2$ are defined as
\begin{equation}\label{21}
\Lambda_1(r_0)= \frac{2(r_0-r_{m1})(r_0-r_{m2})}{1-l}, \qquad \Lambda_1(r_0\to{r_{m2}})=0, 
\end{equation}
\begin{eqnarray}\label{22}
\Lambda_2(r_0)&=&-6DM^2(1-l)+ r_0\Bigg(6M-\frac{r_0}{1-l}\Bigg), \nonumber \\
\Lambda_2(r_0\to{r_{m2}})&=&\frac{1}{2} \left(-8 D+3 \sqrt{9-8 D}+9\right) (1-l) M^2.
\end{eqnarray}

Considering the expansion coefficients contained in Eqs. (\ref{21}) and (\ref{22}) for the strong-field regime $r\to {r_m}$, and analyzing the integration expression of Eq. (\ref{18}), a logarithmic divergence is observed and, therefore, it is necessary to separate this integral into two parts: a regular part $\Delta\phi_R(r_0)$ and a divergent part $\Delta\phi_D(r_0)$. Thus, we have:
\begin{equation}\label{23}
\Delta\phi(r_0) = \Delta\phi_R(r_0) + \Delta\phi_D(r_0).
\end{equation}

Thus, the divergent part of the integration is defined as

\begin{eqnarray}\label{24}
\Delta\phi_D(r_0)&=&\int^1_0\frac{2r_0 \mathrm{d}z}{\sqrt{\Lambda_1(r_0)z+\Lambda_2(r_0)z^2}}= 
-\frac{4r_0}{\sqrt{\Lambda_2(r_0)}}\log\left(\sqrt{\Lambda_1(r_0)}\right) \nonumber \\
&+& \frac{4r_0}{\sqrt{\Lambda_2(r_0)}}\log\left(\sqrt{\Lambda_2(r_0)}\,+\sqrt{\Lambda_1(r_0)+\Lambda_2(r_0)}\right),\qquad \mbox{when} \qquad r_0\to{r_{m2}} \nonumber \\ 
\Delta\phi_D(r_{m2})&=& -\frac{2r_{m2}} {\sqrt{\Lambda_2(r_{m2})}}\log\left(\Lambda_1(r_{m2})\right) +\frac{2r_{m2}}{\sqrt{\Lambda_2(r_{m2})}}\log\left(4\Lambda_2(r_{m2})\right).\nonumber \\ 
\end{eqnarray}

To control the way the integration divergence occurs, we need to expand the coefficient $\Lambda_1(r_0)$ and the impact parameter $\beta(r_0)$ when the radius of the distant photon approaches the outer radius of the photon sphere $r_0{\to}r_{m2}$. Thus, Eqs. (\ref{6C}) and (\ref{13}) become:
\begin{equation}\label{25}
    \beta\left(r_0\right) \simeq \frac{1}{\sqrt{8\eta}}+\frac{\Big(9-8D+3\sqrt{9-8D}\Big)\Big(r_0-r_{m2}\Big)^2}{\sqrt{2}\Big(3+\sqrt{9-8D}\Big)^6\Big(1-l\Big)^5{M^4}\,\eta^{3/2}},
\end{equation} where the parameter $\eta$ is defined as being 
\begin{equation}\label{26}
    \eta=\frac{\Big(3-2D+\sqrt{9-8D}\Big)}{\Big(3+\sqrt{9-8D}\Big)^4{M^2}\Big(1-l\Big)^3}.
\end{equation}

Thus, we can use the expression for the expanded impact parameter above Eq. (\ref{25}) and then substitute it into the expansion coefficient of Eq. (\ref{21}) to consider that this photon, coming from infinity, approaches the outer radius of the photon sphere even further. Therefore, the expansion coefficient in this limit is given by

\begin{eqnarray}\label{27}
\Lambda_1(r_0\to{r_{m2}}) \simeq 6M\sqrt{1-\frac{8D}{9}}\Bigg[\frac{\Big(3+\sqrt{9-8D}\Big)^6{M^4\,\eta\,\Big(1-l\Big)^5}}{2\Big(9-8D+3\sqrt{9-8D}\Big)}\Bigg]^{1/2}\,\times 
\Bigg[\beta\sqrt{\frac{8\Big(3-2D+\sqrt{9-8D}\Big)}{\Big(3+\sqrt{9-8D}\Big)^4\,M^2\,\Big(1-l\Big)^3}}-1\Bigg]^{1/2}.
\end{eqnarray}

Then, substituting equations (\ref{22}), (\ref{25}) and (\ref{27}) into expression (\ref{24}), with some algebraic manipulation, we obtain the divergent part related to the integration of the angular deviation:
\begin{eqnarray}\label{28}
   \Delta\phi_D &=& -\sqrt{\frac{(3 + \sqrt{9 - 8D})(1 - l)}{2 \sqrt{9 - 8D}}}\log\Bigg[\beta\sqrt{\frac{2\left(1 + \sqrt{9 - 8D}\right)}
{\left(3 + \sqrt{9 - 8D}\right)^3 \, M^2 (1 - l)^3}}-1\Bigg]\nonumber\\
    &&+\sqrt{\frac{(3 + \sqrt{9 - 8D})(1 - l)}{2 \sqrt{9 - 8D}}}\log\Bigg[ \frac{8\sqrt{9-8D}}{1+\sqrt{9-8D}}\Bigg] \ .
\end{eqnarray}

The expression constructed above refers to the divergent part of the integration of the angular deviation. Regarding the contribution from the regular part, since this integration cannot be done analytically, we will therefore treat it numerically. Thus, the regular part is given by
\begin{equation}\label{29}
    \Delta\phi_R= \int^{1}_0\,\Bigg(\frac{2\,r_{m2}\,dz}{\sqrt{G(z,r_{m2})}}\Bigg)- \int^{1}_0\,\Bigg(\frac{2\,r_{m2}\,dz}{\sqrt{\Lambda_2(r_{m2})}\,{z}}\Bigg).
\end{equation}

\begin{figure}
    \centering
      \includegraphics[scale=0.41]{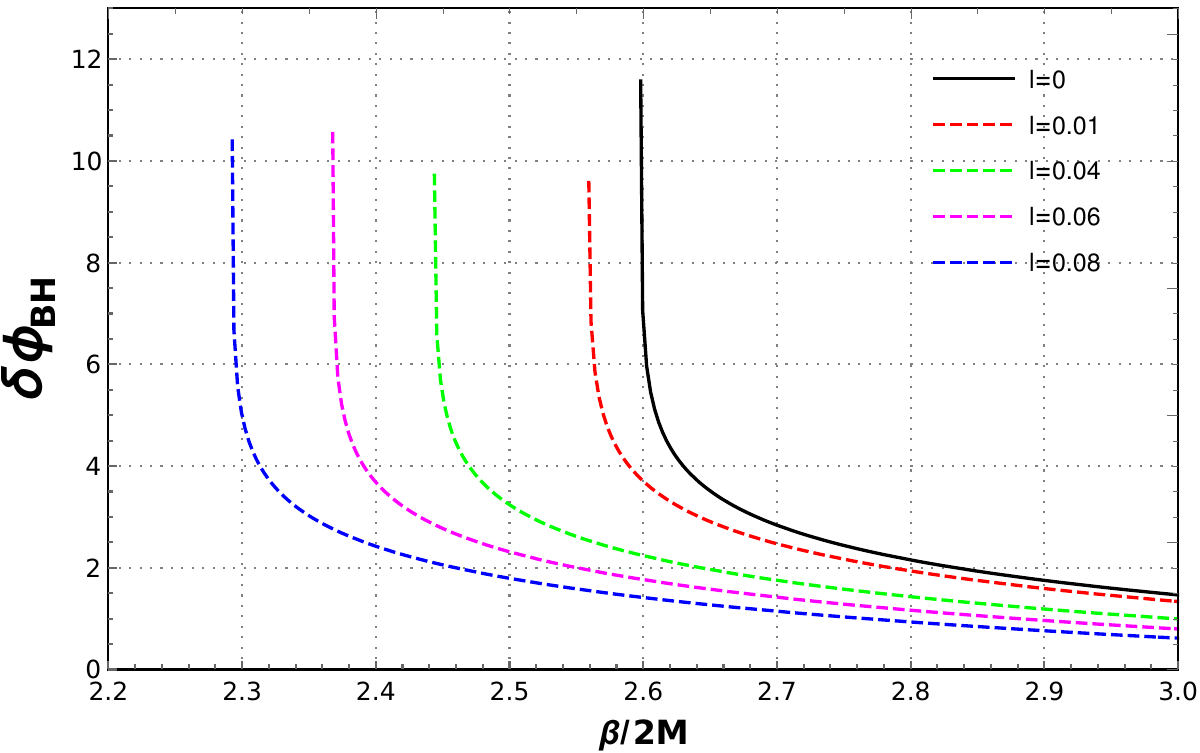}
        \includegraphics[scale=0.41]{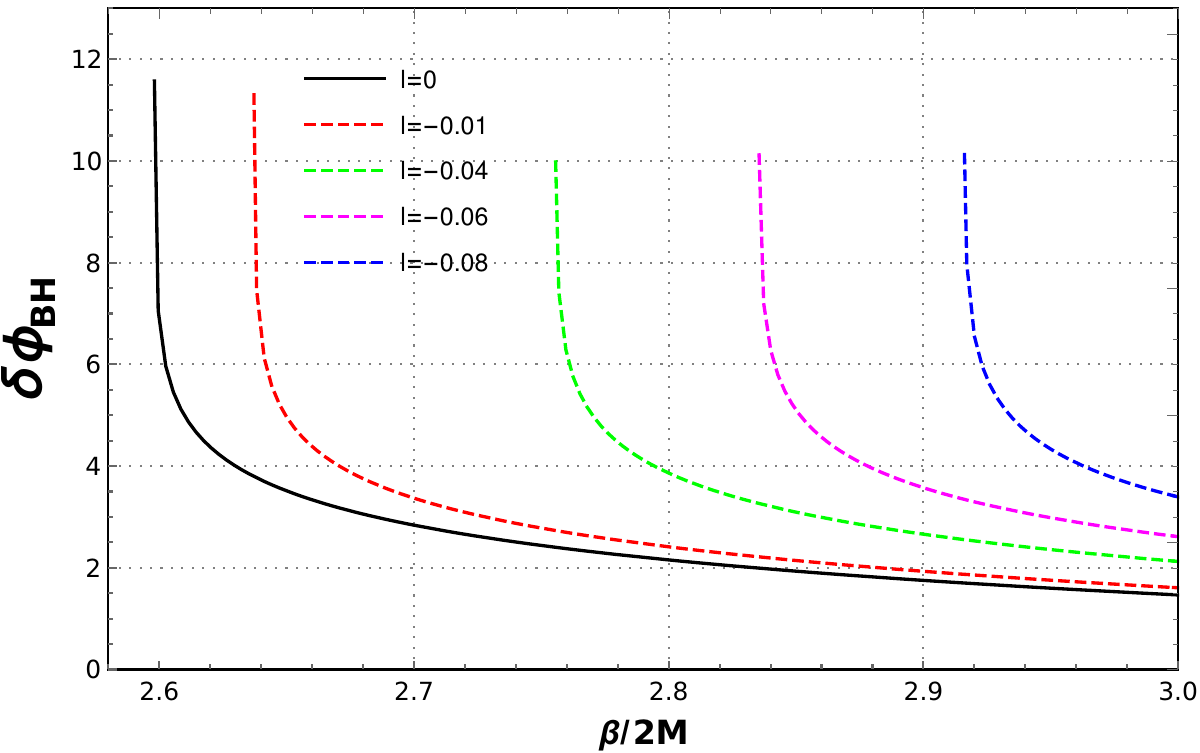}
    \caption{Angular deflection for a Schwarzschild-type BH in Kalb-Ramond gravity.}
    \label{FORTE1}
\end{figure}

Therefore, the total deviation of light due to the presence of the gravitational field in this electrically charged BH in Kalb-Ramond gravity is defined as $\delta\phi=\Delta\phi_D + \Delta\phi_R -\pi$. Since the expression for the regular part of the integration does not have an analytical solution, we will perform the numerical treatment. In a simplified manner, and to verify the effectiveness of the methodology discussed, we can observe in Fig. (\ref{FORTE1}) the representation of the angular deviation suffered by a beam of light for the Schwarzschild BH when subjected to the effects of Kalb-Ramond gravity. Note that in the panel on the left of this figure we consider positive values of the violation parameter, while in the panel on the right we adopt negative values; what changes from one to the other is the intensity of the curvature in the vicinity of the BH, contributing to amplifying or attenuating the effect.

\begin{figure}
    \centering
      \includegraphics[scale=0.41]{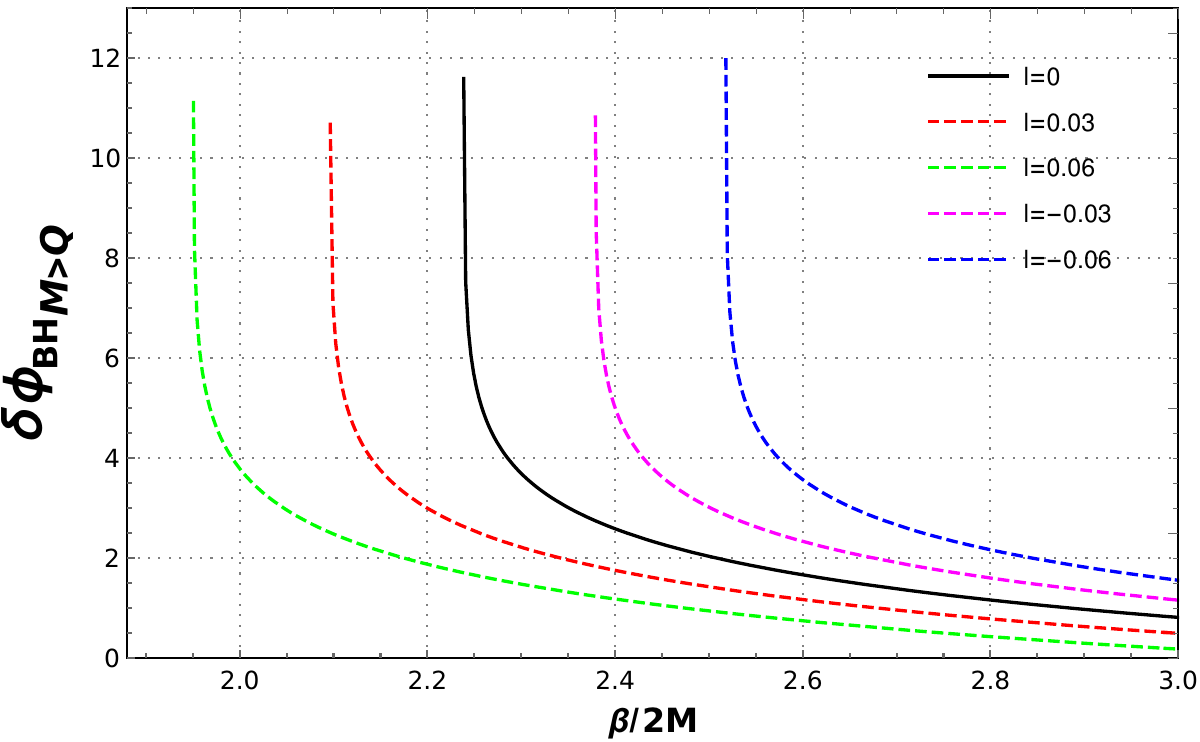}
        \includegraphics[scale=0.41]{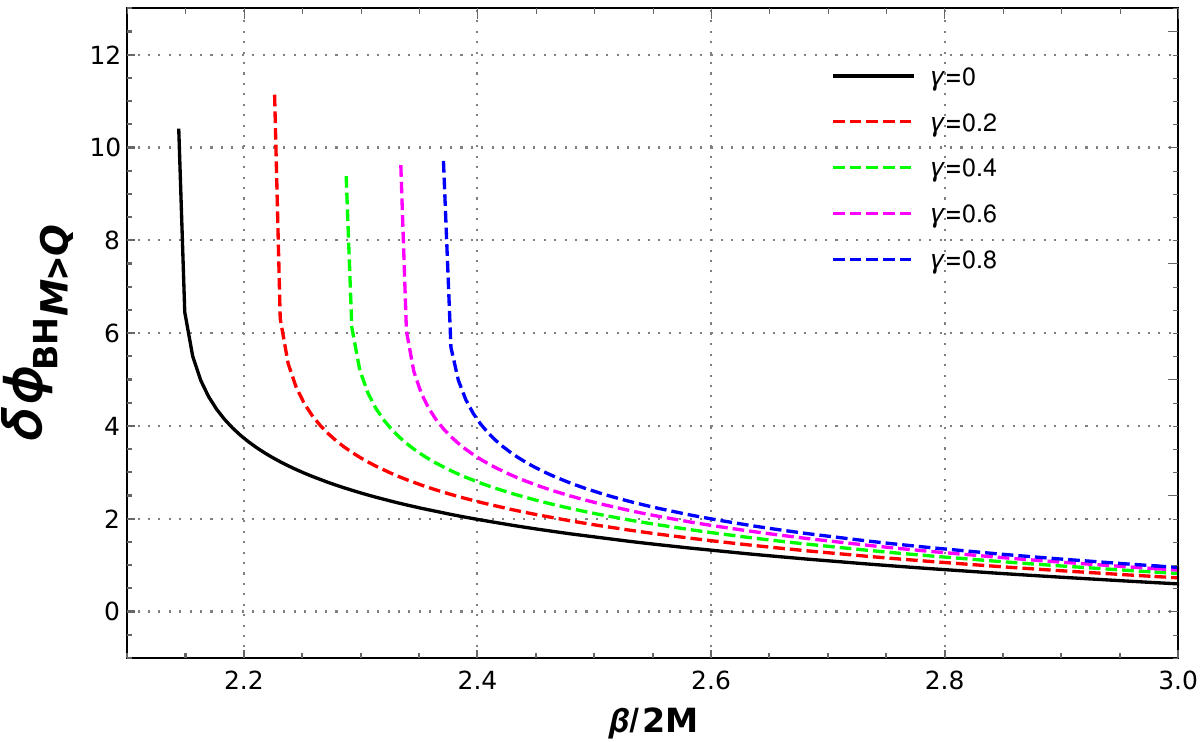}
    \caption{Angular deflection of light for a BH in Kalb-Ramond gravity, where in the panel on the left we fix the parameters $\xi=1$ and $\gamma=0$ and then vary $l$. In the panel on the right we fix $l=0.02$ and $\xi=1$ and then vary the parameter $\gamma$.}
    \label{FORTE2}
\end{figure}

\begin{figure}
    \centering
      \includegraphics[scale=0.41]{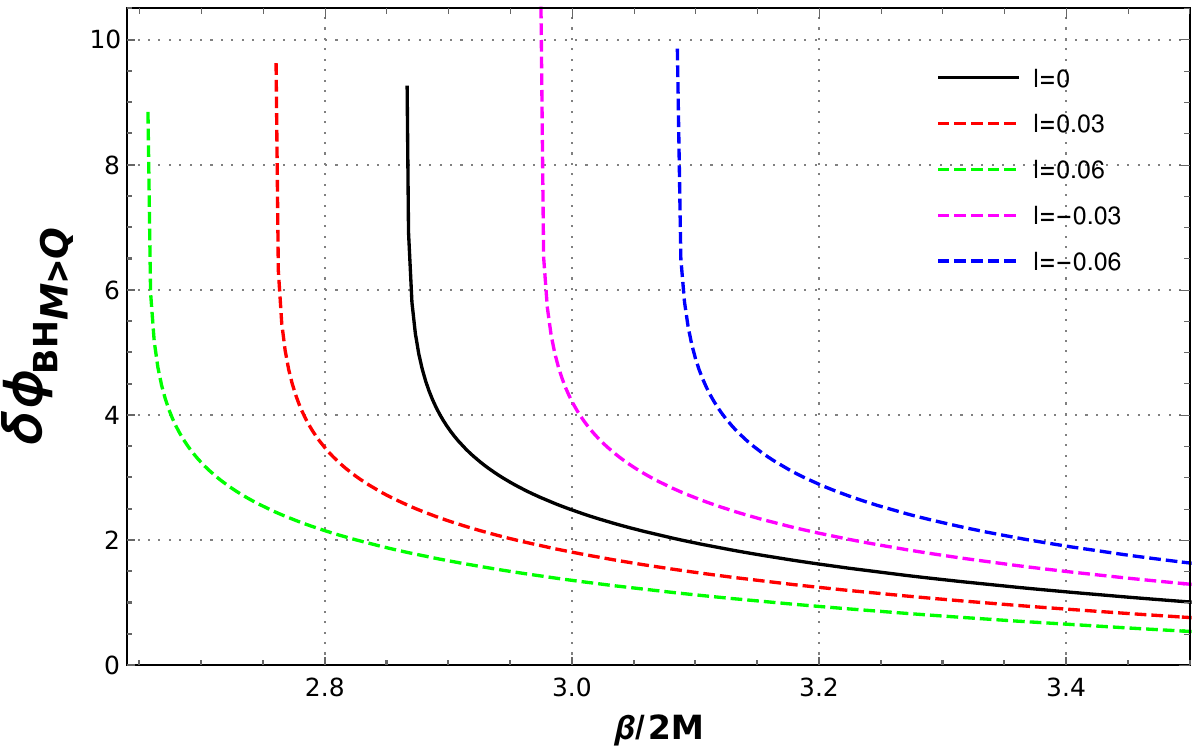}
        \includegraphics[scale=0.41]{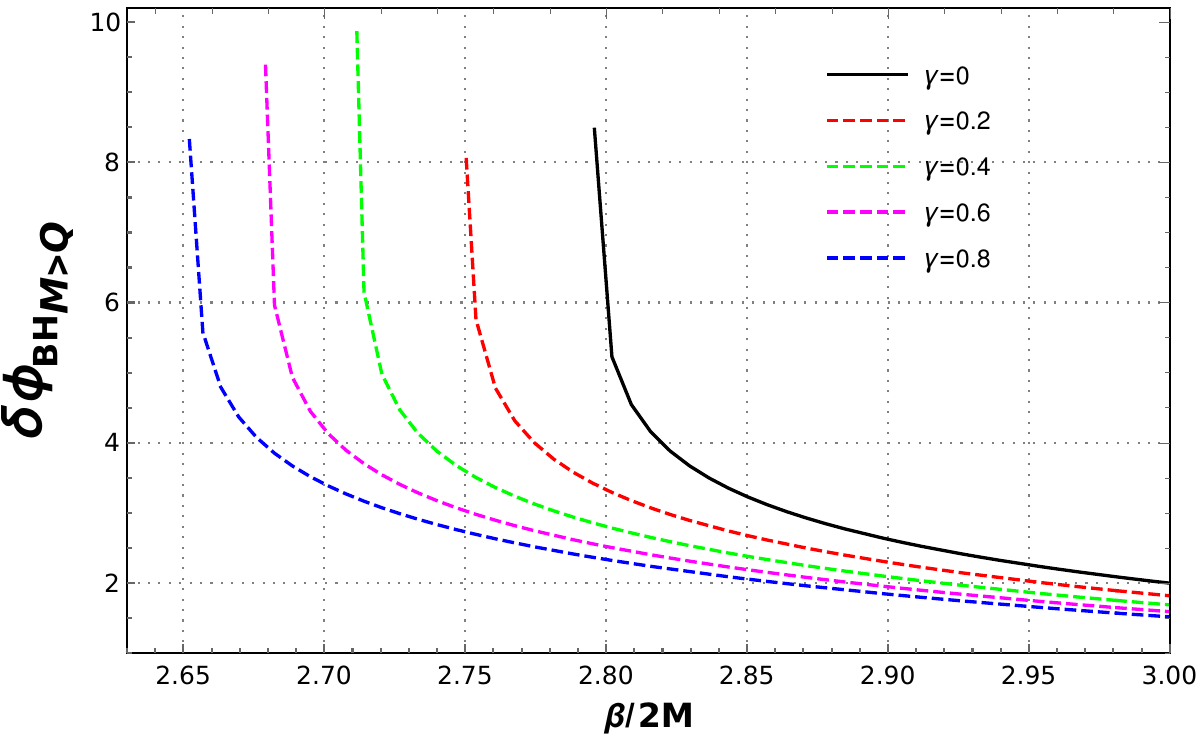}
    \caption{Angular deflection of light for a BH in Kalb-Ramond gravity, where in the panel on the left we fix the parameters $\xi=-1$ and $\gamma=0$ and then vary $l$. In the panel on the right we fix $l=0.02$ and $\xi=-1$ and then vary the parameter $\gamma$.}
    \label{FORTE3}
\end{figure}

In panels of Figs. (\ref{FORTE2}) and (\ref{FORTE3}) we can observe the gravitational deviation for an electrically charged BH subjected to Kalb-Ramond gravity. In both panels, we are considering only the case of the usual BH where mass is greater than the amount of electric charge; for our analysis, we are considering $M=1.2Q$, so that in the first panel (\ref{FORTE2}) we are considering the case where the field is canonical, and the second is the same analysis for the phantom field (\ref{FORTE3}). In particular, we can observe in the panel on the left of the Fig. (\ref{FORTE2}) that we are considering the scenario where the parameter controlling the degree of non-linearity of the electrodynamics is fixed $\gamma=0$ and then we are analyzing the situation when the Lorentz symmetry violation parameter is modifying the light deviation for the case $l=0$ (black continuous curve). Note that we adopt positive and negative values for this parameter, which may indicate amplification or attenuation of the angular deviation. On the other hand, in the panel to the right of the Fig. (\ref{FORTE2}) the continuous black curve represents the angular deviation of the light for the Lorentz symmetry violation parameter set at $l=0.02$ and then we are varying the non-linearity parameter, which in this usual case plays the role of amplifying the angular deviation and in the phantom case (panel to the right of the Fig. (\ref{FORTE3})) it reduces the deviation. Finally, if we compare the panels to the left of Figs. (\ref{FORTE2}) and (\ref{FORTE3}), which show the canonical case and the phantom respectively, what changes from one behavior to the other is only a shift in the coordinate represented by the ratio $\beta/2M$.


\section{Lens equation and observables}\label{sec5}
In this section, we will construct the connection between the angle of deflection of light in the weak-field regime Eq. (\ref{15}), and in the strong-field regime, Eqs. (\ref{28}) and (\ref{29}), with $\delta\phi= \Delta\phi_R + \Delta\phi_D -\pi$, using the master equations of gravitational lensing as a starting point. In this way, we can construct physical quantities that can be theoretically associated with observables. Considering the strong-field regime as a starting point.

\begin{figure}[htb!]
\centering
	{\includegraphics[width=0.75\textwidth]{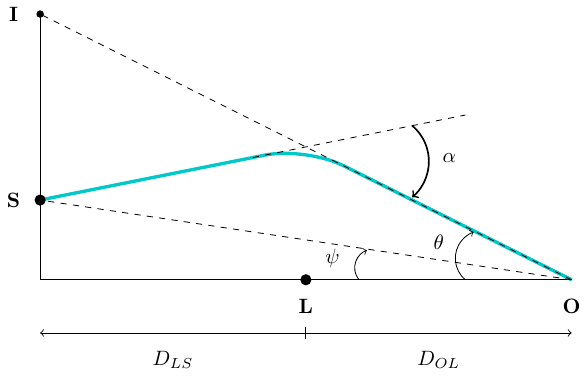}}
    \caption{Light angular deflection diagram.}
\label{LENS1}
\end{figure}

Figure (\ref{LENS1}) illustrates the gravitational lensing diagram. Thus, the light beam emitted by the source \textbf{S} is deflected due to the presence of the BH located at \textbf{L} and subsequently propagates towards the observer \textbf{O}. The angular deviation of the light deflection is denoted by $\alpha$. The angular positions of the source and the image relative to the optical axis, $\overline{LO}$, are represented by $\psi$ and $\theta$, respectively. Thus, considering that the source \textbf{S} is practically aligned with the lens \textbf{L}, a position in which the relativistic images should be more expressive \cite{Bozza:2002af,Vazquez:2003zm}, we then have the lens equation that relates the angular positions $\psi$ and $\theta$ defined as
\begin{equation}\label{LT1}
\psi = \theta -\frac{D_{LS}}{D_{OS}}\Delta\alpha_n,
\end{equation} where $\Delta\alpha_n$ is the deflection angle subtracted from all the loops made by the photons before reaching the observer, that is, $\Delta\alpha_n = \alpha -2n\pi$. In this approach, we use the following approximation for the impact parameter $\beta\approx\theta{D_{OL}}$. Then the contributions to the angular deflection, Eqs. (\ref{28}) and (\ref{29}), are rewritten as:
\begin{equation}\label{LT2}
\alpha(\theta)= -\bar{a}\log\left(\frac{\theta{D_{OL}}}{\beta_c}-1\right) +\bar{b},
\end{equation}
where we have
\begin{eqnarray}\label{LT3}
\beta_c &=&  \sqrt{\frac{\left(3 + \sqrt{9 - 8D}\right)^3 \, M^2 (1 - l)^3}
{2\left(1 + \sqrt{9 - 8D}\right)}}, \qquad \qquad \bar{a}= \sqrt{\frac{(3 + \sqrt{9 - 8D})(1 - l)}{2 \sqrt{9 - 8D}}}, \\\label{LT4}
\bar{b}&=&  \sqrt{\frac{(3 + \sqrt{9 - 8D})(1 - l)}{2 \sqrt{9 - 8D}}}\log\Bigg[ \frac{8\sqrt{9-8D}}{1+\sqrt{9-8D}}\Bigg]   + \Delta\phi_R -\pi, 
\end{eqnarray}
given that the regular part is given by Eq. (\ref{29}). 

To obtain $\Delta\alpha_n$, we expand $\alpha(\theta)$ close to $\theta = \theta^{0}_n$, where $\alpha(\theta^{0}_n)
 = 2n\pi$, and find that
 \begin{equation}\label{LT5}
\Delta\alpha_n= \frac{\partial\alpha}{\partial\theta} \Big|_{\theta=\theta^{0}_n}\left(\theta -\theta^{0}_n\right).
\end{equation}

Evaluating Eq. (\ref{LT2}) on $\theta=\theta^{0}_n$, we have
\begin{equation}\label{LT6}
\theta^{0}_n= \frac{\beta_c}{D_{OL}}(1+e_n), \qquad \mbox{with} \quad e_n= e^{\frac{\bar{b}-2n\pi}{\bar{a}}}.
\end{equation}

Substituting Eq. (\ref{LT2}) and (\ref{LT6}) in Eq. (\ref{LT5}), we have
\begin{equation}\label{LT7}
\Delta\alpha_n= - \frac{\bar{a}D_{OL}}{\beta_c{e_n}}\left(\theta-\theta^{0}_n\right).
\end{equation}

Therefore, substituting Eq. (\ref{LT7}) into Eq. (\ref{LT1}) and performing some algebraic manipulations, we obtain the expression for the angular position of the lens:
\begin{equation}\label{LT8}
\theta_n \approx \theta^{0}_n + \left(\frac{{e_n}\beta_c}{\bar{a}}\right) \frac{D_{OS}(\psi-\theta^{0}_n)}{D_{OL}D_{LS}}.
\end{equation}

Although the deflection of light preserves the surface brightness of the source, the gravitational lensing phenomenon alters the solid angle at which this source is perceived by the observer. Consequently, the total observed flux of an image subject to the gravitational lensing effect is proportional to the magnification factor $\mu_n$, which is defined by $\mu_n=\Big| \frac{\psi}{\theta}\frac{\partial\psi}{\partial\theta}\mid_{\theta=\theta^{0}_n}\Big|^{-1}$. Therefore, substituting Eq. (\ref{LT1}) into Eq. (\ref{LT7}), we have
\begin{equation}\label{LT9}
\mu_n= \frac{e_n(1+e_n)}{\psi\bar{a}}\left(\frac{\beta_c}{D_{OL}}\right)^2\frac{D_{OS}}{D_{LS}}.
\end{equation}

The magnification decreases sharply with increasing $n$, implying that the brightness of the first image, $\theta_1$, predominates over that of subsequent images. Furthermore, the presence of the factor $\left(\frac{\beta_c}{D_{OL}}\right)^2$ indicates that the magnification remains intrinsically small. It is further observed that, in the limit $\psi \to{0}$, corresponding to the maximum alignment between the source, the lens, and the observer, the magnification diverges, resulting in a significant increase in the probability of detecting relativistic images.

\subsection{Observables in the strong-field limit}\label{sec41}

In the previous sections, the positions of the relativistic images, as well as their respective fluxes, were determined as a function of the expansion parameters $\bar{a}$, $\bar{b}$, and $\beta_c$. At this point, we adopt the inverse procedure, reconstructing the expansion coefficients from observational data. This approach allows us to investigate the properties of the object responsible for the gravitational lensing effect and, subsequently, to compare the observational results with the corresponding theoretical predictions. The impact parameter can be expressed in terms of $\theta_\infty$ \cite{B8},
\begin{equation}\label{LT10}
\beta_c= D_{OL}\theta_\infty.
\end{equation}

We will follow Bozza \cite{B8} and assume that only the outermost image $\theta_1$ is resolved as a single image while the others are encapsulated in $\theta_\infty$. Thus, Bozza defined the following observables,
\begin{eqnarray}\label{LT11}
s&=&\theta_1-\theta_\infty=\theta_\infty{e^{\frac{\bar{b}-2\pi}{\bar{a}}}}, \\\label{LT12}
\tilde{r} &=& \frac{\mu_1}{\sum^\infty_{n=2}\mu_n}=e^{\frac{2\pi}{\bar{a}}},
\end{eqnarray} where the parameters $\bar{a}$ and $\bar{b}$ are defined in the expressions (\ref{LT3}) and (\ref{LT4}).

In the previous expressions, $\textbf{s}$ represents the angular separation, while $\tilde{r}$ characterizes the ratio between the flux of the first image and that of the subsequent images. These relationships can be inverted in order to determine the corresponding expansion coefficients. For the analysis of the observables, we assume that the object under consideration has an estimated mass of $4.4\times10^{6}M_{\odot}$ and is located at an approximate distance of $D_{OL}=8.5\text{Kpc}$, values comparable to those associated with the BH at the center of our galaxy \cite{Genzel:2010zy}. In geometric units, we have the rescaling of the mass to $M\to \frac{MG}{c^2}$ and $\theta_\infty=26.5473\mu{arcsecs}$, which is the same used for the Schwarzschild BH.

\begin{figure}
    \centering
    \includegraphics[scale=0.41]{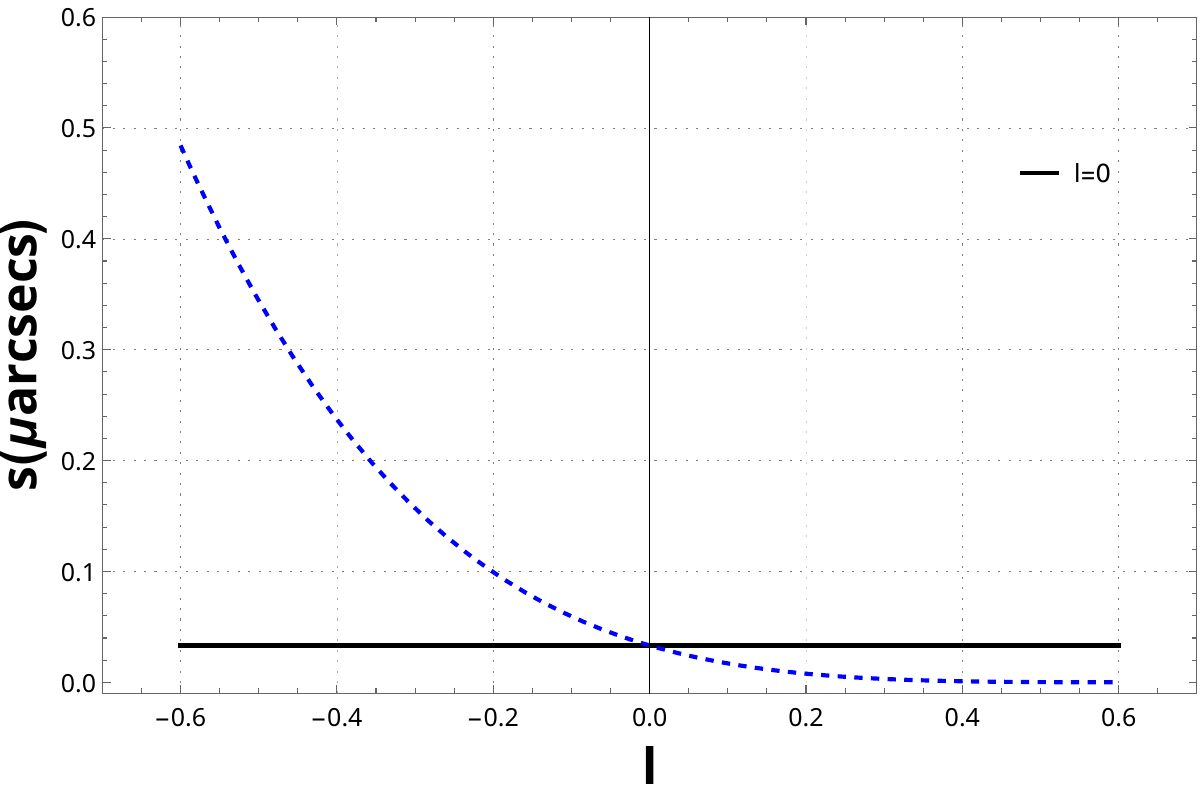}
    \includegraphics[scale=0.41]{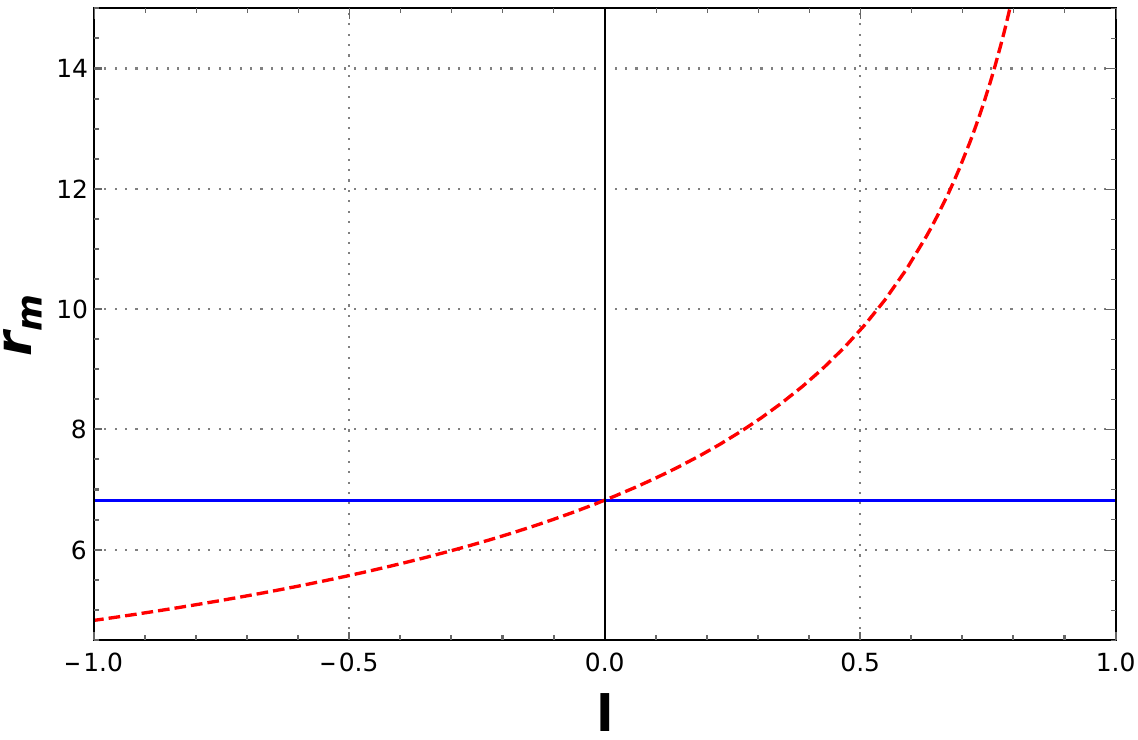}
    \caption{Angular separation and $2.5\,\log_{10}\bar{r}$ as a function of the Lorentz symmetry violation parameter for the Schwarzschild BH.}
    \label{SP1}
\end{figure}

In TABLE \ref{TAB1} we have assigned the variation of the Lorentz symmetry violation parameter to verify how it modifies observables such as the angular separation \textbf{s}, the parameter associated with the flux of the first image \textbf{r} and the angular position at infinity, referring to the Schwarzschild BH and considering the data for the BH at the center of our galaxy and already described in the previous paragraph. The data presented in this table for the angular separation can be visualized in the panel on the left, through the dashed blue curve in the graph of Fig. (\ref{SP1}) and show how this result deviates from the curve referring to the Schwarzschild BH (solid black curve). Note that, as the Lorentz violation parameter approaches $l\to{1}$, the angular separation approaches zero, as we can see in the third column of the table \ref{TAB1}. On the other hand, on the right side of Fig. (\ref{SP1}) we have the graphical representation of what the flux of the first image would be growing as the Lorentz violation parameter tends to $\approx{1}$ (dashed red curve) compared to the Schwarzschild BH (solid black curve). {The comparison with respect to Schwarzschild BH can be observed in table \ref{TAB1} such that the angular separation assumes the approximate value of $s\,\approx {0.033 \, {\mu\,arcsecs}}$ as well as the angular position at infinity tending to $\theta_\infty\, \approx\, 26.55\,{\mu\,arcsecs}$ }

\begin{table}[!ht]
\centering
\caption{Observable quantities as functions of the Lorentz violation parameter $l$ for the Schwarzschild BH.}
\label{Tab:obs1}

\renewcommand{\arraystretch}{1}

\begin{tabular}{c|c|c|c}
\hline\hline
$l$ 
& $\theta_{\infty}\;(\mathrm{\mu\,{arcsecs}})$ 
& $s\;(\mathrm{\mu\,arcsecs})$ 
& $r= 2.5\log_{10}\bar{r}\;(\mathbf{magnitudes})$ \\ 
\hline\hline
-0.6 & 53.7468 & 0.48418 & 5.39317 \\ \hline
-0.4 & 43.8811 & 0.23692  & 5.76554\\ \hline
-0.1 & 34.9096 &  0.09931  & 6.22750\\ \hline
0.0 & 26.5566 & 0.03324 & 6.82188\\ \hline
0.1 & 19.0024  &  0.00782 & 7.62710\\ \hline
0.4 & 12.3424 &  0.00099  & 8.80701\\ \hline
0.6 & 6.71835  & 0.00004 & 10.78630 \\ \hline\hline
\end{tabular} \label{TAB1}
\end{table}

\begin{table}[!ht]
\centering
\caption{Observable quantities as a function of the Lorentz violation parameter $l$ for the electrically charged non-extreme case BH. We fix the parameters $\xi=1$ and $\gamma=0$.}
\label{Tab:obs1A}
\renewcommand{\arraystretch}{1}

\begin{tabular}{c|c|c|c|c|c}
\hline\hline
$l$ 
& $\xi$
& $\gamma$
& $\theta_{\infty}\;(\mathrm{\mu\,{arcsecs}})$ 
& $s\;(\mathrm{\mu\,arcsecs})$ 
& $r= 2.5\log_{10}\bar{r}\;(\mathbf{magnitudes})$ \\ 
\hline\hline
-0.10 & 1 & 0 & 27.6257 & 0.08850 & 5.98069 \\ \hline
-0.06 & 1 & 0 & 25.7383 & 0.07708  & 5.99791\\ \hline
-0.02 & 1 & 0 & 23.8403 &  0.06832  & 5.98658\\ \hline
 0.0 & 1 & 0 &  22.8829 & 0.06502 & 5.96435 \\ \hline
0.02 & 1 & 0 &  21.9163  &  0.06257 & 5.92588\\ \hline
0.06 & 1 & 0 &  19.9385 &  0.06119  & 5.77078 \\ \hline
0.10 & 1 & 0 &  17.8468  & 0.06938 & 5.39397\\ \hline\hline
\end{tabular} \label{TAB1A}
\end{table}

\begin{table}[!ht]
\centering
\caption{Observable quantities as a function of the Lorentz violation parameter $l$ for the electrically charged non-extreme case BH. We fix the parameters $\xi=-1$ and $\gamma=0$.}
\label{Tab:obs1B}
\renewcommand{\arraystretch}{1}

\begin{tabular}{c|c|c|c|c|c}
\hline\hline
$l$ 
& $\xi$
& $\gamma$
& $\theta_{\infty}\;(\mathrm{\mu\,{arcsecs}})$ 
& $s\;(\mathrm{\mu\,arcsecs})$ 
& $r= 2.5\log_{10}\bar{r}\;(\mathbf{magnitudes})$ \\ 
\hline\hline
-0.10 & -1 & 0 & 33.0769 & 0.04733 & 6.80626 \\ \hline
-0.06 & -1 & 0 & 31.5372 & 0.03675  & 6.92208 \\ \hline
-0.02 & -1 & 0 & 30.0365 &  0.02807  & 7.12969 \\ \hline
 0.0 & -1 & 0 &  29.3012 & 0.02437 & 7.21835 \\ \hline
0.02 & -1 & 0 &  28.5761  &  0.02105 & 7.31051 \\ \hline
0.06 & -1 & 0 &  27.1572 &  0.01547  & 7.50619 \\ \hline
0.10 & -1 & 0 &  25.7812  & 0.01111 & 7.71864 \\ \hline\hline
\end{tabular} \label{TAB1B}
\end{table}

\begin{figure}
    \centering
    \includegraphics[scale=0.41]{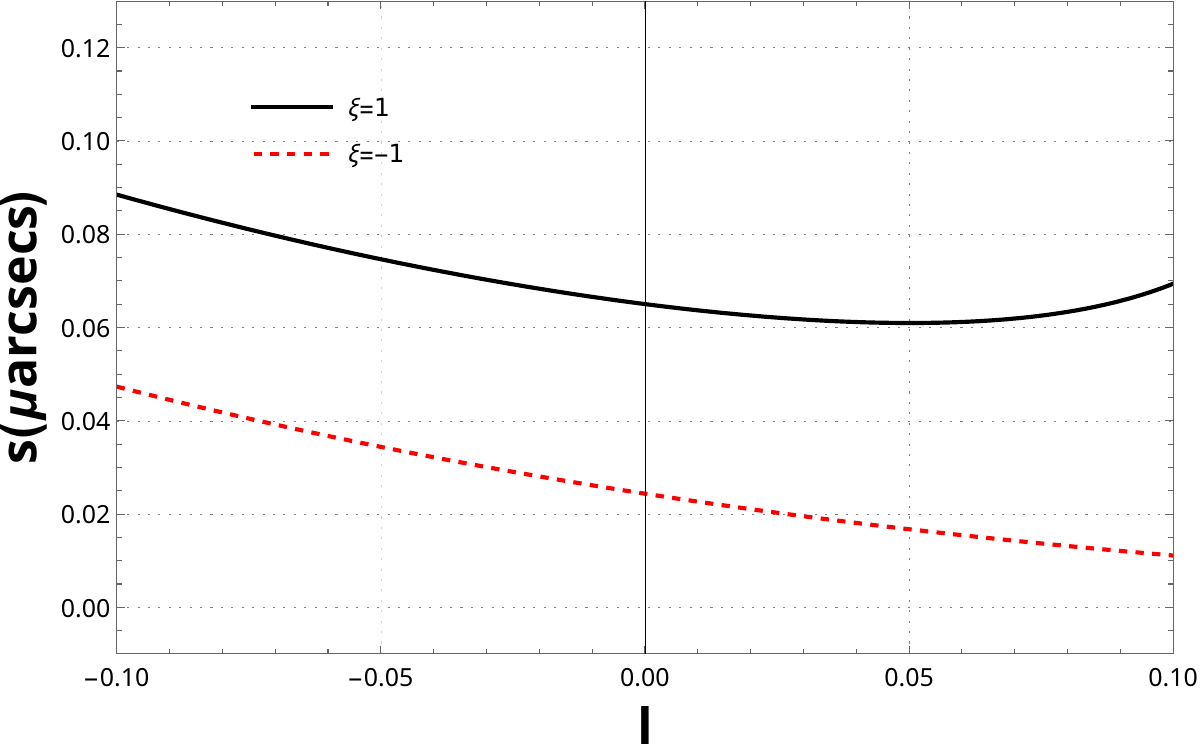}
    \includegraphics[scale=0.41]{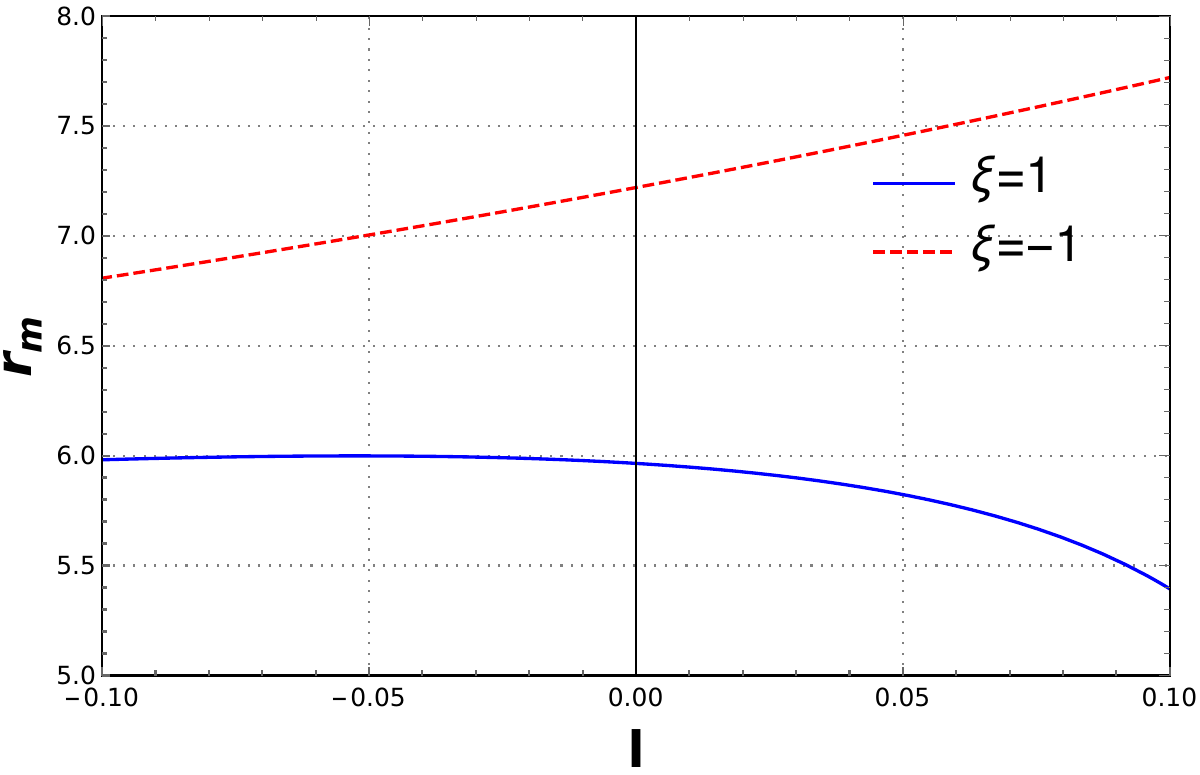}
    \caption{Angular separation and $2.5\,\log_{10}\bar{r}$ as a function of the Lorentz symmetry violation parameter for the electrically charged non-extreme BH.}
    \label{SP24}
\end{figure}

In tables \ref{TAB1A} and \ref{TAB1B} we construct data for some observables for the non-extreme BH $M=1.2\,Q$, taking into account the BH data at the center of our galaxy and, respectively, considering the case where the field is canonical $\xi=1$ and the phantom  case $\xi=-1$. In this way, we kept the parameter that regulates nonlinearity fixed $\gamma=0$ and then varied the Lorentz symmetry violation parameter. 

For illustrative purposes, the data contained in these tables for angular separation and the relationship of the flow of the first image with the others can be visualized through the panels found in the Fig. (\ref{SP24}). Thus, note in the continuous black curve (on the left) that the angular separation for the canonical field case reaches a minimum point and then increases again, a behavior that is consistent with table \ref{TAB1A}. However, for the phantom case (red dotted curve on the left), the angular separation continues to decrease as the Lorentz violation parameter approaches $l\,\approx{0.1}$ and this behavior is in accordance with the table \ref{TAB1B}. In the panel to the right of the Fig. (\ref{SP24}), we can observe that in the continuous blue curve, the flux ratio of the first image to the others reaches a maximum value and then begins to decrease as the Lorentz symmetry violation parameter approaches $l\,\approx{0.1}$. This behavior is consistent with the data contained in table \ref{TAB1A}. On the other hand, this same quantity for the phantom case (dashed red curve) shows an increasing behavior in this interval and can be observed in table \ref{TAB1B}.

\begin{table}[!ht]
\centering
\caption{Observable quantities as a function of the parameter controlling the nonlinearity $\gamma$ for the non-extreme case of an electrically charged BH. We fix the parameters $\xi=1$ and $l=0.2$.}
\label{Tab:obs1C}
\renewcommand{\arraystretch}{1}

\begin{tabular}{c|c|c|c|c|c}
\hline\hline
$l$ 
& $\xi$
& $\gamma$
& $\theta_{\infty}\;(\mathrm{\mu\,{arcsecs}})$ 
& $s\;(\mathrm{\mu\,arcsecs})$ 
& $r= 2.5\log_{10}\bar{r}\;(\mathbf{magnitudes})$ \\ 
\hline\hline
0.2 & 1 & 0.2 & 13.6365 & 0.10856 & 3.44546 \\ \hline
0.2 & 1 & 0.4 & 15.2317 & 0.03112  & 5.95292 \\ \hline
0.2 & 1 & 0.6 & 16.1322 &  0.01953  & 6.54114 \\ \hline
0.2 & 1 & 0.8 & 16.7607 & 0.01508 & 6.85321 \\ \hline
0.2 & 1 & 0.9 & 17.0102  & 0.01376 & 6.96180 \\ \hline
0.2 & 1 & 0.95 & 17.1224 &  0.01323  & 7.00826 \\ \hline
0.2 & 1 & 0.98 &  17.1862  & 0.01295 & 7.03405 \\ \hline\hline
\end{tabular} \label{TAB1C}
\end{table}

\begin{table}[!ht]
\centering
\caption{Observable quantities as a function of the parameter controlling the nonlinearity $\gamma$ for the non-extreme case of an electrically charged BH. We fix the parameters $\xi=-1$ and $l=0.2$.}
\label{Tab:obs1D}
\renewcommand{\arraystretch}{1}

\begin{tabular}{c|c|c|c|c|c}
\hline\hline
$l$ 
& $\xi$
& $\gamma$
& $\theta_{\infty}\;(\mathrm{\mu\,{arcsecs}})$ 
& $s\;(\mathrm{\mu\,arcsecs})$ 
& $r= 2.5\log_{10}\bar{r}\;(\mathbf{magnitudes})$ \\ 
\hline\hline
0.2 & -1 & 0.2 & 21.9794 & 0.00462 & 8.24997 \\ \hline
0.2 & -1 & 0.4 & 21.5011 & 0.00495  & 8.16849 \\ \hline
0.2 & -1 & 0.6 &  21.0930 &  0.00527  & 8.09435 \\ \hline
0.2 & -1 & 0.8 & 20.7466 & 0.00557 & 8.02769 \\ \hline
0.2 & -1 & 0.9 & 20.5940  & 0.00571 & 7.99716 \\ \hline
0.2 & -1 & 0.95 & 20.5224 &  0.00578  & 7.98258 \\ \hline
0.2 & -1 & 0.98 & 20.4809  & 0.00582 & 7.97405 \\ \hline\hline
\end{tabular} \label{TAB1D}
\end{table}

\begin{figure}
    \centering
    \includegraphics[scale=0.41]{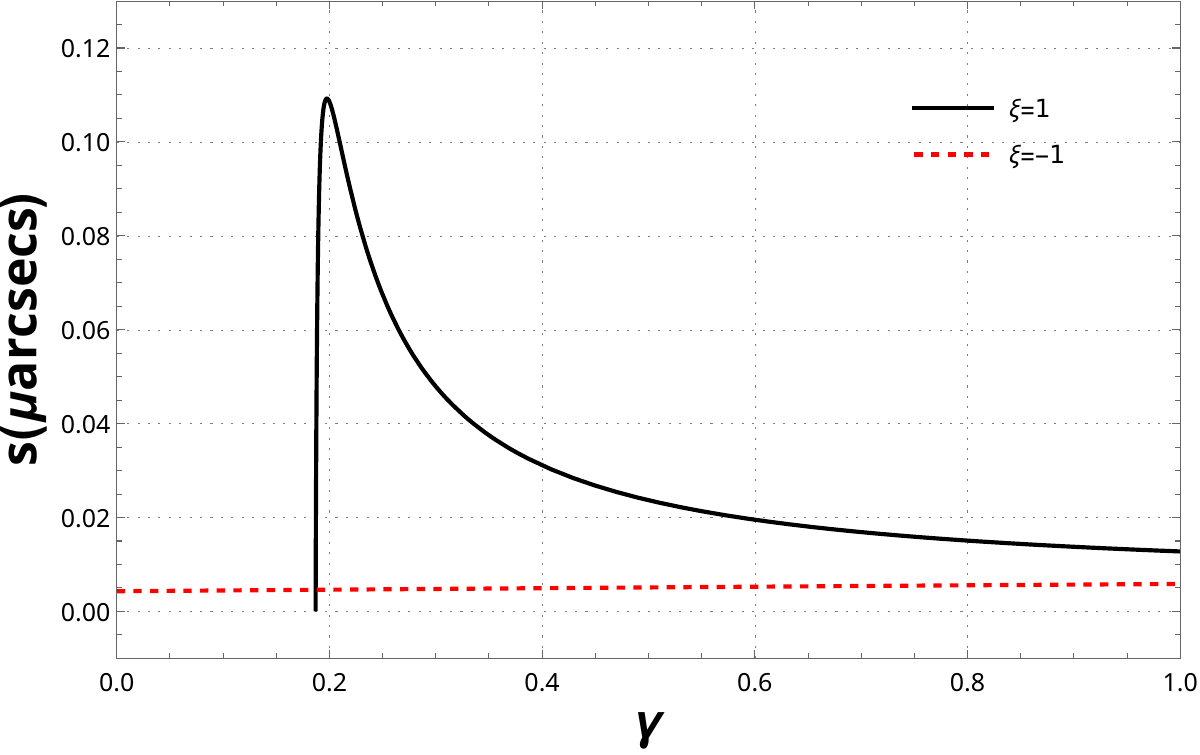}
    \includegraphics[scale=0.41]{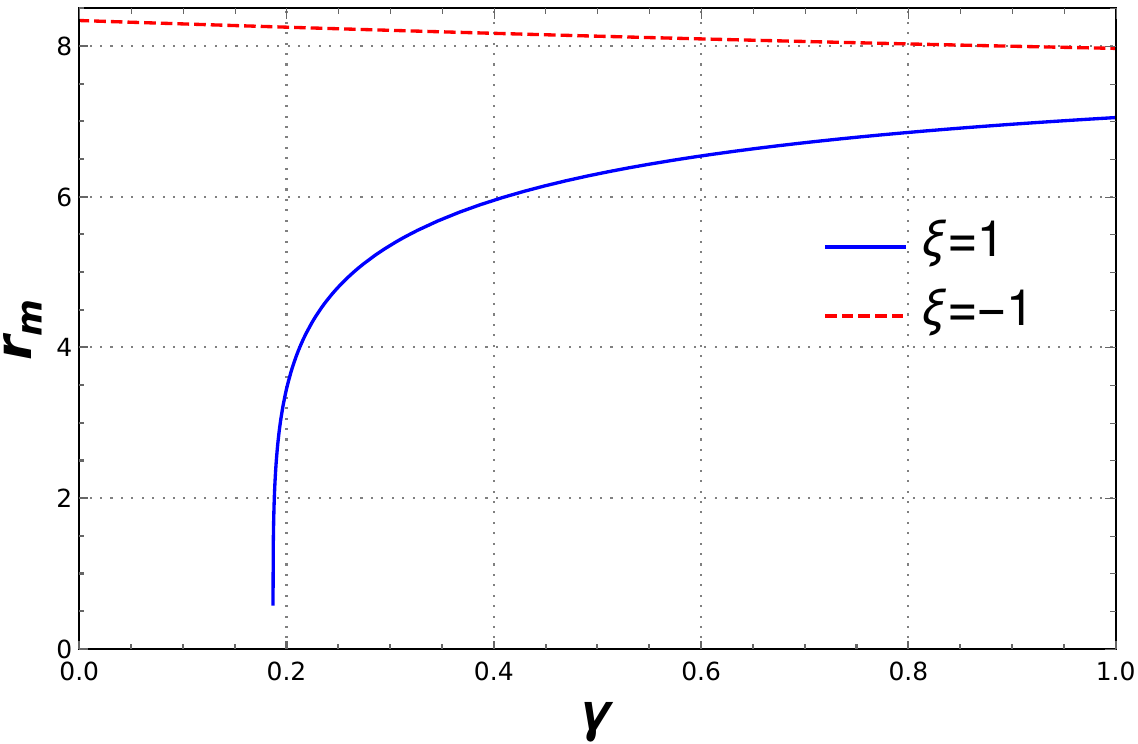}
    \caption{Angular separation and $2.5\,\log_{10}\bar{r}$ as a function of the parameter $\gamma$ for the electrically charged non-extreme BH.}
    \label{SP35}
\end{figure}

In tables \ref{TAB1C} and \ref{TAB1D}, we constructed data for some observables related to the non-extreme BH $M=1.2\,Q$, taking into account the BH data at the center of our galaxy and, respectively, considering the case where the field is canonical $\xi=1$ and the phantom case $\xi\,=-1$. In this way, we fixed the Lorentz symmetry violation parameter $l\,=0.2$ and then varied the parameter that controls the nonlinearity. We have created a graphical representation of how varying the parameter that controls nonlinearity is modifying the angular separation and magnification associated with the image flow, which can be seen in Fig. (\ref{SP35}).

{In the red dotted curve of the panel to the left Fig. (\ref{SP35}), the angular separation assumes slightly increasing values for the phantom case, as the nonlinearity parameter tends to $\gamma\to{1}$, in accordance with table \ref{TAB1D}. In the case where the field is canonical (black continuous curve), the angular separation reaches a maximum point at $\gamma\,\approx0.187$, and then this quantity begins to have an imaginary contribution that does not appear in the table \ref{TAB1C}, since the smallest value chosen was $\gamma=\,0.2$.}

{In the panel to the right of Fig. (\ref{SP35}), we have the curve that relates the flux in the first image to the others for the phantom field, and we observe that it decreases as the nonlinearity parameter tends to $\gamma\,\to{1}$, a behavior that is consistent with the data presented in table \ref{TAB1D} (see the dotted red curve). On the other hand, for the canonical case (continuous blue curve), this quantity tends to increase as the parameter approaches $\gamma\,\to{1}$ and is consistent with the table \ref{TAB1C}. However, for the value of the parameter $\gamma\,\approx0.187$, this quantity, as well as the angular separation (continuous black curve on the left), begins to have an imaginary contribution.}



\subsection{Observables in the weak-field limit}\label{sec42}

The next step is to analyze the observables in the weak-field regime, which means that the impact parameter is very large relative to the object's mass, $\beta\gg{M}$, so that the light beam does not form loops. We can also note that for simplicity we will develop this analysis considering only a Schwarzschild-type BH contained within Kalb-Ramond gravity, so as to consider only the linear term in the mass $M$. Therefore, the expression Eq. (\ref{13}) becomes
\begin{equation}\label{LTW1}
\delta\phi \sim  \pi\Big(-1+\sqrt{1-l}\Big) + \frac{4M(1-l)^{2}}{\beta}.
\end{equation}

Taking as a starting point the equation that relates the angular positions of the source and the image, Eq. (\ref{LT1}), we analyze the particular case of perfect alignment between the source, the compact object, and the observer, a situation that corresponds to the limit $\psi = 0$. In this way, we have to
\begin{equation}\label{LTW2}
\theta= \frac{D_{LS}}{D_{OS}}\Delta\alpha_n,
\end{equation} where $\Delta\alpha_n$ is given by Eq. (\ref{LTW1}). Therefore, substituting Eq. (\ref{LTW1}) into Eq. (\ref{LTW2}) and keeping only terms up to first order in the mass $M$, we obtain a second-degree algebraic equation, which when solved gives us the angular position $\theta_E$ for the Einstein ring:
\begin{eqnarray}\label{LTW3}
\theta&=&\theta_E= \frac{D_{LS}\,\pi\Big(-1+\sqrt{1-l}\Big)}{2\,D_{OS}}+ \frac{1}{2}\Bigg[\frac{D_{LS}}{D_{OS}}\Bigg(\frac{D_{LS}}{D_{OS}}\Big(\pi\Big(-1+\sqrt{1-l}\Big)\Big)^2 +\frac{16\,M\Big(1-l\Big)^{2}}{D_{OL}}\Bigg)\Bigg]^{1/2}.
\end{eqnarray}

Thus, in the expression above we can clearly observe that the Lorentz symmetry violation parameter, as well as the mass of the Schwarzschild BH, is modifying the relative position of the Einstein ring. In this sense, we can also point out that, as observed in the light deflection in both weak- and strong-field regimes, the position of the Einstein ring varies depending on whether we are considering the violation parameter with positive or negative values. From the expression above regarding the angular position of the Einstein ring, we can consistently recover the expression referring to the Schwarzschild BH when the effects of Lorentz symmetry violation are deactivated $l\to{0}$ \cite{Soares:2023uup}:
\begin{equation}\label{LTW31}
\theta=\theta_E= \sqrt{\frac{4MD_{LS}}{D_{OS}D_{OL}}}. 
\end{equation}

We can also calculate the radius of the Einstein ring from the approximation $\beta\approx\theta{D_{OL}}$ and using Eq. (\ref{LTW3}) \cite{Soares:2023uup,Han:1996hb}. Therefore, we have
\begin{equation}\label{LTW4}
R_E= D_{OL}\theta_E.
\end{equation} 

Therefore, observables $R_E$ and $\theta_E$ can be determined by considering different values of the parameter $l$, associated with the violation of the Lorentz symmetry, as well as the mass $M$. Within the scope of this analysis, restricted to the weak-field regime, we investigate the gravitational lensing effect, taking as an example a star located in the galactic bulge \cite{Abe:2010ap}. To this end, we consider the following values for the parameters $D_{OL}=4Kpc$ and $D_{OS}=8Kpc$.

\begin{table}[!ht]
\centering
\caption{Radius of the Einstein ring and angular position as a function of the Lorentz violation parameter $l$.}
\label{Tab:obs2}

\renewcommand{\arraystretch}{1}

\begin{tabular}{c|c|c}
\hline\hline
$l$ 
& $R_E\;(\mathrm{km})$ 
& $\theta_E\;(\mathrm{arcsecs})$ \\ 
\hline\hline
-0.6 & $5.14\times10^{16}$ & 85831.20 \\ \hline
-0.4 & $3.55\times10^{16}$ & 59362.00 \\ \hline
-0.1 & $9.46\times10^{15}$ & 15814.1 \\ \hline
0.0 & $1.27\times10^{12}$ & 2.12 \\ \hline
0.1 & $1.31\times10^{8}$  & $2.2\times10^{-4}$ \\ \hline
0.4 & $1.32\times10^{7}$ &  $2.2\times10^{-5}$  \\ \hline
0.6 & $3.60\times10^{6}$  & $6.02\times10^{-6}$  \\ \hline\hline
\end{tabular} \label{TAB2}
\end{table}

In table \ref{TAB2} we have collected numerical data for both the angular position and the radius of the Einstein ring, taking into account the Schwarzschild BH which is deformed by the Lorentz symmetry violation parameter $l$. Thus, note that as the Lorentz symmetry violation parameter increases $l\,\approx\,{1}$, both the angular position and the radius of the Einstein ring decrease. {As noted in table \ref{TAB2}, an increase of approximately four orders of magnitude in $R_E$ and $\theta_E$ is observed when passing from $l=0$ to $l=-0.1$. This behavior expresses a clear transition of the dominant regime in the spacetime geometry. In the Schwarzschild limit ($l=0$), lensing is a purely local effect governed by the smallness of the mass ratio $M/D_{OL}$. However, for any $l \neq 0$, the background topological term asymptotically associated with a topological charge comes into play, corresponding to the first term in Eq.~(\ref{LTW3}). Since the contribution of this topological effect is much greater than the local gravitational contribution (the term that depends on the mass $M$), it completely dominates the weak-field scenario, drastically expanding the observables.}

{In the weak-field limit with $l=0$, the spacetime reduces to the Schwarzschild solution, thereby preserving the same observables. Gravitational lensing in this regime was also investigated in \cite{Soares:2023uup,Soares:2025hpy}, considering a loop quantum gravity-motivated model based on a bulge star, similar to the present work. Upon vanishing the parameters that characterize loop quantum gravity, the model reduces to the Schwarzschild solution, with which our results are in agreement.}






\section{Conclusion}\label{sec6}

In this work, we initially establish the general relationships and conserved quantities, as well as the geodesic trajectories, for an electrically charged beam subjected to Kalb-Ramond gravity that incorporates the Lorentz symmetry violation parameter. Subsequently, we focus on investigating the gravitational deflection of a light beam when subjected to the weak-field regime. To verify the consistency of the methodology employed, we obtained the gravitational shift for the most general background investigated, and then we found the angular shift for a more simplified background Eq. (\ref{15}), which would be the Schwarzschild BH in the Kalb-Ramond theory Eq. (\ref{16}), as well as the shift referring only to the Schwarzschild BH when considering that the Lorentz symmetry violation effects were turned off $l\,\to{0}$ Eq. (\ref{17}). We constructed graphical representations to analyze the relationship between mass and electric charge parameters, encompassing the configurations of BH $M=Q$, $M>Q$, and $M<Q$. Furthermore, we also considered the modifications created by the Lorentz violation parameter $l$, the degree of nonlinearity of the electrodynamics $\gamma$, and the canonical and phantom configurations $\xi\,=\pm{1}$.

In the second part of the work, we performed the light deflection calculation for the strong-field regime in order to analytically obtain the expressions for the expansion coefficients, aiming to control the logarithmic divergence that occurred in the limit where the impact parameter converges to the critical radius of the photon sphere and thus obtain the final expression for the angular deviation Eq. (\ref{29}). From the equations obtained for the weak-field regime, we constructed the gravitational lensing expressions so that we could correlate these theoretical calculations with the observables associated with relativistic images. Due to technical difficulties, for the calculation of the angular position and radius of the Einstein ring, we considered only the expression for the deviation of light in spacetime from Kalb-Ramond Eq. (\ref{16}) and considered only first-order terms in the mass $M$. For the strong-field regime, data relating to the BH at the center of our galaxy were used to calculate the angular deflection \cite{Genzel:2010zy}, and then we constructed some tables and graphs by varying the parameters, and then we can observe how these alter the observables. With regard to observables in the field regime, we use data from a bulge star \cite{Abe:2010ap} to verify how the radius of the Einstein ring and the angular position are modified with the variation of the Lorentz symmetry violation parameter.


\section*{Acknowledgments}
\hspace{0.5cm} 

The authors express their gratitude for the fruitful discussions held with Professor Manuel  E. Rodrigues. A. A. Araújo Filho is supported by Conselho Nacional de Desenvolvimento Cient\'{\i}fico e Tecnol\'{o}gico (CNPq) and Fundação de Apoio à Pesquisa do Estado da Paraíba (FAPESQ), project numbers 150223/2025-0 and 1951/2025. M. V. de S. Silva is supported by CNPq/PDE 200218/2025-5. R. L. L. Vit\'oria is supported by Conselho Nacional de Desenvolvimento Cient\'{\i}fico e Tecnol\'{o}gico (CNPq) and by Universidade Estadual do Marnh\~ao (UEMA), projects numbers 150420/2025-0 and by the EDITAL N. 102/2025-PPG/CPG/UEMA, respectively.


	\bibliography{main}
	\bibliographystyle{unsrt}
	
\end{document}